\documentclass[fleqn,usenatbib]{mnras}
\usepackage[T1]{fontenc}
\usepackage{ae,aecompl}
\usepackage{array}
\usepackage{graphicx}	% Including figure files
\usepackage{amsmath}	% Advanced maths commands
\usepackage{amssymb}	% Extra maths symbols
\title[Transients from ONe WD -- NS/BH Mergers]{Transients from ONe White Dwarf -- Neutron Star/Black Hole Mergers}

\author[Bobrick et al.]{
Alexey Bobrick$^{1}$\thanks{E-mail: alexey@astro.lu.se},
Yossef Zenati$^{2,3}$, 
Hagai B. Perets$^{2}$,
Melvyn B. Davies$^{1,4}$,
Ross Church$^{1}$
\\
$^{1}$Lund Observatory, Department of Astronomy and Theoretical physics, Box 43, SE 221-00 Lund, Sweden\\
$^{2}$Physics Department, Technion - Israel Institute of Technology, Haifa 3200004, Israel\\
$^{3}$Physics and Astronomy Department, Johns Hopkins University, Baltimore, MD 21218, USA\\
$^{4}$Centre for Mathematical Sciences, Lund University, Box 118, SE 221-00 Lund, Sweden\\
}

\date{Accepted XXX. Received YYY; in original form ZZZ}
 
\pubyear{2021}

\hypersetup{draft}

\usepackage{todonotes}

\begin{document}
\label{firstpage}
\pagerange{\pageref{firstpage}--\pageref{lastpage}}
\maketitle

\begin{abstract}
We conduct the first 3D hydrodynamic simulations of oxygen-neon white dwarf-neutron star/black hole mergers (ONe WD-NS/BH mergers). Such mergers constitute a significant fraction, and may even dominate, the inspiral rates of all WD-NS binaries. We post-process our simulations to obtain the nuclear evolution of these systems and couple the results to a supernova spectral synthesis code to obtain the first light curves and spectra for these transients. We find that the amount of $^{56}$Ni synthesised in these mergers grows as a strong function of the WD mass, reaching typically $0.05$ and up to $0.1\,{\rm M}_\odot$ per merger. Photodisintegration leads to similar amounts of $^4$He and about a ten times smaller amount of $^1$H. The nuclear yields from these mergers, in particular those of $^{55}$Mn, may contribute significantly to Galactic chemical evolution. The transients expected from ONe WD-NS mergers are dominantly red/infrared, evolve on month-long timescales and reach bolometric magnitudes of up to -16.5. The current surveys must have already detected these transients or are, alternatively, putting strong constraints on merger scenarios. The properties of the expected transients from WD-NS mergers best agree with faint type Iax supernovae. The Vera Rubin Observatory (LSST) will be detecting up to thousands of merging ONe WD-NS systems per year. We simulate a subset of our models with 2D axisymmetric FLASH code to investigate why they have been challenging for previous studies. We find that the likely main challenge has been effectively modelling the nuclear statistical equilibrium regime in such mergers.
\end{abstract}


%%%%%%%%%%%%%%%%%%%%%%%%%%%%%%%%%%%%%%%%%%%%%%%%%%

%%%%%%%%%%%%%%%%% BODY OF PAPER %%%%%%%%%%%%%%%%%%

%\section*{Color legend}

%\todo[inline]{To-do; title of the block (authors A, B,.. responsible)}
%\turnintotext[inline]{Turn into normal text}
%%\review[inline]{Review (author A, B,.. who has reviewed the block)}
%\done[inline]{Done}

\section{Introduction}
\label{sec:Intro}

\subsection{ONe WD-NS/BH binaries in the transient sky}

%\review[inline]{Peculiar supernovae are trending (Alexey)}
The recent advent of wide-field all-sky synoptic surveys, e.g. PTF \citep{Law2009,Rau2009}, Pan-STARRS \citep{Chambers2016}, ZTF \citep{Bellm2019}, ATLAS \citep{Smith2020}, among others, as well as the improvement in rapid spectroscopic follow-up capabilities, have brought about a variety of new types of optical transients. More than a dozen new classes have been discovered so far \citep{Milisavljevic2018}, including Ca-rich transients \citep{Perets2010,Kasliwal12}, type Iax-events \citep{Jha2006,Phillips2007}, 1991bg-like events \citep{Leibundgut1993}, kilonovae \citep{Cowperthwaite2017} and others. Compared to standard classes of supernovae, these transients show a broad range of peak magnitudes, lightcurves, spectra, and  multimessenger observables, including radio, X-ray, gamma-ray, neutrino and gravitational wave signals, e.g. \citet{Abbott2017,Abbott2017b}.

Given that binary and single stars may end their lives through a variety of energetic events, e.g. \citet{Hurley2002, Belczynski2008}, peculiar transients provide a new and exciting window into stellar evolution. Identifying peculiar supernovae with their progenitor systems allows us to study the progenitor populations, the physics of high-energy events, their environments and, indirectly, the populations of systems which do not end their lives violently. Many peculiar transients still lack a conclusive identification. The Vera Rubin observatory \citep{LSST2009}, expected to become fully operational in 2022, is expected to observe thousands of new transient events per night \citep{Ridgway2014}.

%\review[inline]{CO WD-NS/ONe WD-NS transients are about to be found (Alexey)}
Mergers of white dwarfs (WDs) and neutron stars (NSs) are a common outcome of binary stellar evolution \citep{Nelemans2001}. However, these events are still lacking a conclusive optical counterpart. According to existing models, e.g. \citet{Zenati19b}, CO WD-NS mergers likely produce faint (bolometric magnitudes up to $\approx -15$) rapidly-evolving red transients, potentially detectable by current surveys. Mergers of ONe WDs with NSs, due to their more massive and more compact WDs and higher merger energies, may potentially lead to even brighter events. However, transients from ONe WD-NS mergers have not been modelled yet in the literature. Here we model such mergers, study their observational signatures, and then provide observational predictions and compare them with observations. 

Based on the rapidly-evolving lightcurve, large O, Mg and Si content and high inferred $~^{54}{\rm Fe}/^{56}{\rm Ni}$ ratio, \citet{McBrien2019} and \citet{Gillanders2020} argued that SN AT2018kzr might be the product of an ONe WD-NS or BH merger. If correct, SN AT2018kzr might be the first system to connect theoretical models of ONe WD-NS/BH mergers, and WD-NS mergers generally, to observations.

\subsection{ONe WD-NS/BH binary populations}

%\review[inline]{How WD-NS/BH binaries are observed (Alexey)}
WD-NS binaries show several observational signatures. Detached systems may be observed as binary radio pulsars, e.g. \citet{Lorimer2008}, and will be observable as gravitational wave (GW) sources, e.g. by LISA \citep{Tauris2018}. The binaries with sufficiently short orbital periods of less than about $10\,\rm{h}$ spiral into contact due to gravitational wave emission \citep{Peters1964}. Upon coming into contact, binaries with a He WD of less than approximately $\approx0.2\,{\rm M}_\odot$ will survive the onset of mass transfer \citep{Bobrick2017} and evolve into long-living ultra-compact X-ray binaries, UCXBs \citep{Savonije1986}, after a short phase as ultra-luminous X-ray sources, ULXs \citep{Bildsten2004}. The WD-NS binaries with He WDs of mass above $\approx0.2\,{\rm M}_\odot$ or with CO or ONe WDs merge shortly after coming into contact \citep{Bobrick2017} and may be observed as optical transients \citep{Metzger2012,Zenati19b}. Similarly, WD-BH binaries may also be observed as GW sources when detached. Insipralling WD-BH binaries will result in mergers only when they contain a massive ONe WD and a low-mass, $5$~--~$6\,{\rm M}_\odot$, BH \citep{Church2017}.

%\review[inline]{Observations constrain inspiral rates (Alexey)}
Observed binary radio pulsars can constrain the inspiral rates of WD-NS binaries. There are, currently, $10$ binary pulsars with a WD companion of more than $0.08\,{\rm M}_\odot$ known that will spiral into contact within Hubble time \citep[ATNF catalogue]{Manchester2005}. Among these,  J1141-6545 \citep{Kaspi2000} contains the youngest, non-recycled, pulsar with a massive WD ($M_{\rm WD}\gtrsim 0.98\,{\rm M}_\odot$). It strongly dominates the Galactic merger rate \citep{Kim2004,Kalogera2005}, with its estimated rate being $10$~--~$100\,{\rm Myr}^{-1}$ \citep{OShaughnessy2010}. The empirical merger rate for the whole population is estimated to be about $260\,{\rm Myr}^{-1}$ \citep{Bobrick2017}.

%\review[inline]{Constraints from pop.synthesis (Alexey)}
Pulsar J1141-6545 represents a population of WD-NS binaries where the NS formed {\it second} and the WD companion is relatively massive, e.g. \citet{PortegiesZwart1999,Davies2002,Church2006}. Recent population synthesis of NS-WD binaries by \citet{Toonen2018} predicts that most inspiralling systems have WD masses between $0.7$ and $1.3\,{\rm M}_\odot$ and Galactic merger rates of $100-200\,{\rm Myr}^{-1}$, in agreement with the observed empirical rates. In comparison, at the present day, $10^{-3}$ type Ia supernovae occur in the Galaxy per one solar mass of stars formed, e.g. \citet{Maoz2017}. Assuming a Galactic star-formation rate of \citet{Licquia2015}, this corresponds to the Galactic type Ia rate of $1650\,{\rm Myr}^{-1}$. Therefore, most merging WD-NS mergers contain a relatively massive WD and produce transient events at a rate of $6$~--~$20\,\%$ that of type Ia supernovae. For merging WD-BH binaries, in comparison, there are no empirical or population-based constraints available at the moment.

\subsection{Existing simulations of WD-NS/BH binary mergers}

%\review[inline]{Hydro simulations of WD-NS mergers and transients: Before the merger (Alexey)}
WD-NS mergers and the resulting transients have now been simulated at almost all of the stages of their evolution. Following the gravitational inspiral, WD-NS binaries start transferring material at low rates which gradually increase over time. \citet{Bobrick2017} modelled this phase with $3{\rm D}$ hydrodynamic simulations. They found that WD-NS binaries subsequently merge if $M_{\rm WD}>0.2\,{\rm M}_\odot$ (or survive otherwise). For merging systems, the mass transfer rates keep increasing for hours (for massive WDs) to years (for lower-mass WDs) until the mass transfer rates become so large that the WD gets tidally shredded into a disk over several orbital periods.

%\review[inline]{Hydro simulations of WD-NS mergers and transients. During and right after the merger. (Alexey)}
Most of the nuclear elements get synthesised during the time of the WD being shredded into a disk. This phase lasts between about $100$ and $1000$ seconds, depending on the WD mass. The first nuclear simulations of WD-NS mergers represented the WD as a one-dimensional static \citep{Metzger2012} or time-dependent \citep{Margalit16} disk. These works modelled He and CO WD-NS mergers, as well as white dwarf-black hole (WD-BH) mergers, and showed that such mergers produce moderate amounts of $^{56}{\rm Ni}$, of a few times $0.01\,{\rm M}_{\odot}$, and hence lead to relatively faint transients. \citet{Margalit17} also used time-dependent $1{\rm D}$ models to simulate the long-term evolution of the disk after the merger and argued that the remnants of the WD might lead to planet formation. Subsequently, more accurate $2{\rm D}$ simulations with an Eulerian FLASH code \citep{Fryxell2000} approximated the merger phase by representing the WD as an axisymmetric torus spreading viscously into a disk \citep{Fernandez2013, Zenati19a, Fernandez19}. 
\citet{Fernandez2013} modelled He and CO WD-NS binaries and focused on the effects of choosing the nuclear network and the possibility of detonation. \citet{Zenati19a} provided an independent estimation of the nuclear yields with a larger nuclear network, made use of self-consistent accretion disks, accounted for the self-gravity of the disk, and modelled CO and HeCO WD-NS mergers. \citet{Fernandez19} simulated CO WD-NS and ONe WD-BH mergers, focusing on the structure of the outflows. Recently, \citet{Zenati19b} simulated CO WD-NS mergers in $3{\rm D}$, starting with realistic initial conditions. They used the Lagrangian Smoothed Particle Hydrodynamics (SPH) code Water, post-processed with a detailed nuclear network, and compared their results to $2{\rm D}$ FLASH simulations, showing that realistic initial conditions lead to nuclear yields generally compatible, within about 1 dex, with the earlier $2{\rm D}$ modelling. \citet{Zenati19b} also applied a supernova spectral synthesis code to their results, producing the first multi-band lightcurves (LCs) and spectra for CO WD-NS mergers. General-relativistic, non-nuclear, aspects of WD-NS mergers were studied by \citet{Paschalidis2011}, although with rescaled versions of WD models.

As follows from \citet{Zenati19b} and as we describe further in the text, $3{\rm D}$ models capture the full $3{\rm D}$ process of nucleosynthesis in WD-NS binaries that occurs over several orbital periods during the WD disruption, as the WD material circularizes into a disk. In comparison, $2{\rm D}$ simulations describe idealized nucleosynthesis happening over fractions of a period due to a converging axisymmetric flow from a WD represented by a toroid. On the other hand, the existing 2D models in FLASH self-consistently follow the nuclear energy input (though only from a 19 elements network) during the simulation, and then make use of post-processing with a large network, while the existing 3D models only implement nucleosynthesis in a post-processing step. Furthermore, the existing 2D models have a higher resolution (5--8 km) at the innermost part of the accretion disk than the current 3D models (70 km), and potentially better capture the dynamics and gravitational energy transfer in these regions. Subsequently, to some extent the two approaches complement each other with each having some advantages and disadvantages, and together help bracket the possible range of energetics and nucleosynthesis products. We highlight the differences between the two methods in more detail further.

%\review[inline]{What do we do (Alexey)}
All the existing studies of WD-NS mergers only modelled the binaries with WD masses of up to $0.8\,{\rm M}_\odot$. The WD-NS binaries with more massive WDs significantly heat up and start nuclear burning before the WD material circularises, which presents a problem for axisymmetric $1{\rm D}$ or $2{\rm D}$  simulations, e.g. \citet{Metzger2012, Fernandez2013, Zenati19a}. At the same time, the Galactic observations, as well as population synthesis studies, suggest that most merging systems have WD masses in the range between $0.7\,{\rm M}_\odot$ and $1.3\,{\rm M}_\odot$, e.g. \citet{Toonen2018}. Similarly, the first WD-NS/BH transient candidate, SN AT2018kzr, was also argued to originate from disruption of an ONe WD by an NS or a BH \citep{Gillanders2020}. In this study, we perform $3{\rm D}$ hydrodynamic simulations with nuclear processing of massive CO/ONe WD-NS/BH binaries with WD masses of $0.9\,{\rm M}_\odot$ and larger to cover this yet unexplored domain. We also produce the first light curves and spectra for these mergers and compare them to all possible currently observed classes of transient events. Based on the comparison, we show that ONe WD-NS mergers most likely produce faint type Iax supernovae, while an ONe WD-BH merger was likely responsible for SN AT2018kzr. In addition, we perform $2{\rm D}$ FLASH simulations without nuclear evolution for a subset of these systems and further comment on why they have been challenging to model with earlier codes.

\begin{figure}
\includegraphics[width=\linewidth]{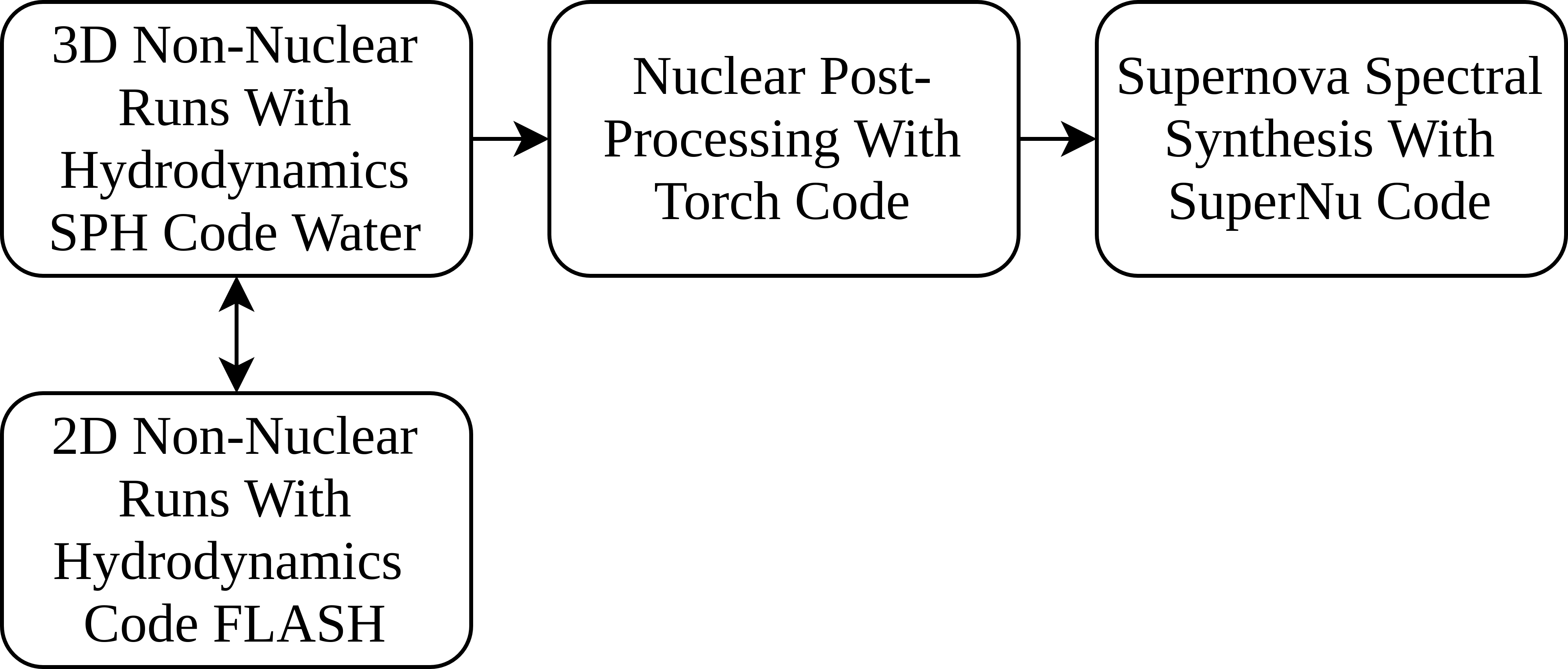}
\caption{Summary of the codes we used in this study and the relations between them. We performed the simulations of WD-NS/BH mergers with non-nuclear 3D hydrodynamics Lagrangian SPH code Water. The resulting models were post-processed with nuclear processing code Torch. We then used the SPH Water code models and the nuclear yields from the Torch code to construct supernova lightcurves and spectra with the supernova spectral synthesis code SuperNu. Finally, we compared our 3D SPH simulations with non-nuclear runs from 2D Eulerian code FLASH to gain insight into why massive ONe WD-NS binaries have been challenging to model in previous 2D studies.}
\label{fig:CodeStructure}
\end{figure}

\begin{table}
\begin{tabular}{ |c|c|c|c|c|} 
 \hline
 Model & $M_{\rm WD},\,{\rm M}_\odot$ & $M_{\rm NS/BH},\,{\rm M}_\odot$ & Comp. & Accretor \\ 
 \hline
 U & $0.9$ & $1.4$ & ONe & NS \\ 
 U$_{2}$ & $0.9$ & $1.4$ & CO & NS \\  
 V & $1.0$ & $1.4$ & ONe & NS  \\
 W & $1.1$ & $1.4$ & ONe & NS  \\
 X  & $1.2$ & $1.4$ & ONe & NS  \\ 
 X$_{2}$ & $1.2$ & $2.0$ & ONe & NS  \\ 
 X$_{3}$ & $1.2$ & $5.0$ & ONe & BH  \\
 Y  & $1.3$ & $1.4$ & ONe & NS  \\ 
 \hline
\end{tabular}
\caption{The WD-NS/BH models simulated in this study. The columns show the Model Name, the mass of the WD donor ($M_{\rm WD}$), the mass of the NS/BH accretor ($M_{\rm NS/BH},\,{\rm M}_\odot$), the WD composition, and the accretor type. CO and ONe compositions correspond to homogeneous $X_{\rm C}=X_{\rm O}=0.5$ and $X_{\rm O}=0.75$, $X_{\rm Ne}=0.25$ mixtures, respectively.}
\label{tab:Models}
\end{table}

\section{Numerical Simulations}

%\review[inline]{One-paragraph summary of the section (Alexey)}
In this section, we describe how we model the WD-NS/BH mergers and obtain the lightcurves and spectra for the transients they produce. In Figure~\ref{fig:CodeStructure}, we summarise the relations between the codes we use. First, we simulate the mergers with the $3{\rm D}$ hydrodynamics Lagrangian SPH code Water \citep{Bobrick2017}, starting from realistic initial conditions. Further, we post-process the simulations with the nuclear-processing code Torch \citep{Timmes1999, Timmes2000b} in order to obtain the yields for the nuclear elements, including $^{56}{\rm Ni}$, synthesised during the mergers. Subsequently, we use these yields to calculate the light curves and spectra with the supernova spectral synthesis code SuperNu \citep{Wollaeger2013, Wollaeger2014}. Furthermore, we simulate some of the mergers with non-nuclear axisymmetric 2D hydrodynamic simulations using an Eulerian code FLASH \citep{Fryxell2000}. We compare the results of the SPH and FLASH code to gain insight into why massive WD-NS mergers have been challenging with earlier 2D studies. In the following sections, we provide the details on each of the steps in the modelling. We use the methods similar to \citet{Zenati19b} and only summarise our model's main components here.

\subsection{The studied WD-NS/BH models}

%\review[inline]{What models do we simulate (Alexey)}
We summarise the WD-NS binaries we model in Table~\ref{tab:Models}. We focus on the binaries containing massive CO and ONe WDs, which have not been modelled in earlier studies. For all the CO and ONe WD models, we use $X_{\rm C}=X_{\rm O}=0.5$ and $X_{\rm O}=0.75$, $X_{\rm Ne}=0.25$ compositions, respectively, as suggested by stellar evolution models, e.g. \citet{GarciaBerro1997}. Models U and U$_2$ represent binaries with $0.9\,{\rm M}_\odot$ WDs with, correspondingly, ONe and CO compositions and a $1.4\,{\rm M}_{\odot}$ NS. While ONe WDs likely do not exist at such low masses, e.g. \citet{Schwab2019}, we use these two models to study how the composition of the WD affects the resulting transients. Models V, W, X and Y contain a $1.4\,{\rm M}_{\odot}$ NS accretor and ONe WD donors with masses of $1.0$, $1.1$, $1.2$ and $1.3\,{\rm M}_\odot$. These models represent the population of WD-NS binaries similar to PSR J1141-6545, which are the dominant population of WD-NS binaries that spiral into contact in the Galaxy \citep{Kim2004,Kalogera2005}. We also consider model X$_2$ which contains a $1.2\,{\rm M}_\odot$ ONe WD donor and a massive $2.0\,{\rm M}_\odot$ NS accretor. We chose such NS mass to study the effect of a heavier accretor, and it also covers the possible mass range of NS companions to WDs in the Galaxy, e.g. \citet{Lattimer2012}. Finally, we also consider model X$_3$ which contains a $1.2\,{\rm M}_\odot$ ONe WD donor and a $5.0\,{\rm M}_\odot$ BH accretor to model the transients resulting from WD-BH mergers. While small compared to the BH masses observed in the Galactic BH X-ray binaries \citep{CorralSantana2016,Tetarenko2016}, the BH mass is chosen based on the requirement that the WD-BH mass transfer results in a merger rather than a surviving stably-transferring X-ray binary \citep{Bobrick2017,Church2017}. We also use model X$_3$ to compare our results to the recent work by \citet{Fernandez19}, who modelled mergers of $1.2\,{\rm M}_\odot$ ONe WDs with $5\,{\rm M}_\odot$ BHs with 2D axisymmetric simulations with the Eulerian FLASH code. Since massive white dwarfs are not expected to contain a helium layer on their surface \citep{Zenati2018}, we focus entirely on helium-free models here.

\subsection{3D hydrodynamic simulations with the SPH code}

\label{sec:MethodSPHSetup}

%\review[inline]{SPH code introduction and motivation (Alexey)}
We perform hydrodynamic simulations of ONe WD-NS/BH mergers with a $3\,{\rm D}$ Lagrangian SPH code Water \citep{Bobrick2017}, similarly to the approach we used for CO WD-NS mergers \citep{Zenati19b}. The Water code uses the most recent SPH prescriptions available and has been used as a part of the Oil-on-Water code to model the onset of mass transfer in WD-NS binaries \citep{Bobrick2017}, see also \citet{Church2009}. Compared to earlier axisymmetric 2D simulations of WD-NS mergers, the $3{\rm D}$ code simulations start from realistic initial conditions. With a $3{\rm D}$ code, we can resolve the donor and the disk's geometry and model the gradual growth of the mass transfer rate preceding the actual merger. These are also the first simulations of ONe WD-NS mergers since massive WDs have proven challenging to simulate with 2D codes, as discussed, e.g. in \citet{Fernandez19}. Therefore, we also analyse why such massive WDs are possible to model with our code, in contrast to the 2D codes. We model the nuclear evolution in a separate post-processing step since nuclear evolution is dynamically unimportant for these systems, e.g. \citet{Zenati19a}. We have also verified that nuclear burning in our simulations is also relatively unimportant dynamically. The small dynamical role of nuclear evolution may be understood from the local nuclear energy released being several times smaller than the local gravitational binding energy.

%\review[inline]{SPH code components (Alexey)}
We solve the Navier-Stokes equations for hydrodynamics by discretizing the Lagrangian following \citet{Springel2010}. Such discretization ensures exact energy and momentum conservation, only limited by the accuracy of the integrator and the gravity solver. For the integrator, we use the kick-drift-kick (KDK) scheme, as described in \citet{Quinn1997,Springel2005,Cullen2010}, and we implement the gravity tree following \citet{Benz1990}. We limit the maximum radius of SPH particles in a Lagrangian and continuous way as done in \citet{Bobrick2017}, partially mitigating the so-called fall-back problem when ejected particles become large and acquire too many neighbours upon falling back on the donor. We use Wendland W6 kernels \citep{Dehnen2012}, which prevent pairing instability by construction, and set the number of the neighbours to $400$. We use artificial viscosity from \citet{Cullen2010}. This prescription switches on the viscosity only near shocks and does not damp sound waves. We base the artificial conductivity on the prescription from \citet{Bobrick2017, Hu2014}, which reduces discontinuities in thermal energy near shocks. Finally, we use a time step limiter by \citet{Saitoh2009}, which improves shock treatment. We refer to \citet{Bobrick2017} for full implementation details of the code.

%\review[inline]{Single model setup and relaxation (Alexey)}

We construct the WD models based on stellar structure profiles obtained by solving stellar structure equations, similarly to \citet{Bobrick2017}, for models shown in Table~\ref{tab:Models}. We initialised all CO WDs with homogeneous $X_{\rm C}=X_{\rm O}=0.5$ compositions, and all ONe WDs were assigned homogeneous $X_{\rm O}=0.75$, $X_{\rm Ne}=0.25$ compositions. The stellar models follow the Helmholtz equation of state (EOS) from \citet{Timmes2000}, which describes a mixture of ions, partly degenerate electron gas and radiation in local thermodynamical equilibrium with the matter. All the models were assigned initial temperatures of $10^5\,{\rm K}$, corresponding to cold perfectly degenerate WDs. We then constructed the SPH models of the WDs by mapping SPH particles onto appropriately spaced spherical shells \citep{Saff1997, Bobrick2017,Raskin2016} and making them follow the stellar profiles. For all the models we used a resolution of $100\,{\rm K}$ SPH particles. We used SPH models with equal particle masses to avoid numerical artefacts reported in earlier studies, e.g. \citet{LorenAguilar2009,Dan2011}. In the SPH simulations, we used the same Helmholtz EOS as for the stellar models. Finally, we set up the NS/BH accretors as point masses with their gravity softened by the Wendland W6 kernel with a core radius of $0.1\,R_{\rm WD}$.

%\review[inline]{Binary setup, and simulations (Alexey)}

We first relaxed the single WD models in isolation over six dynamical times. Then we placed the donors in a binary with the NS/BH at $1.8\,a_{\rm RLOF}$, where $a_{\rm RLOF}$ is the Roche lobe overflow separation and relaxed the binaries in a co-rotating frame, continually spiralling in the donor down to $0.975\,a_{\rm RLOF}$ over two orbital periods. Then we relaxed the binary at the final separation over an additional half a period, so that the main simulation starts with an existing stable mass-transfer stream, see \citet{Bobrick2017}. We removed the small number of particles accreted onto the NS during the relaxation stage. Subsequently, we ran the simulations for $20$ orbital periods, which fully covers the WD disruption phase and the formation of a fully developed disk.

\subsection{2D hydrodynamic simulations with the FLASH code}

%\review[inline]{FLASH simulations for comparing 2D to 3D (Alexey)}
We also simulate the ONe WD-NS/BH mergers for models U, U$_2$, X$_2$ and X$_3$ in $2{\rm D}$ with the publicly available FLASH code \citep{Fryxell2000} without nuclear burning. We use these runs to compare how the commonly used $2{\rm D}$ approach to modelling such mergers compares to the simulations with the $3{\rm D}$ SPH code. In these $2{\rm D}$ simulations, we initialise the WD as a torus and viscously spread it into a disk, which is intended to resemble the disk from the tidally shredded WDs in the actual realistic WD-NS mergers. As we discuss in \citet{Zenati19b}, the $2{\rm D}$ simulations represent the discs only qualitatively since one cannot fully represent the disruption energetics and geometry of 3-dimensional WD-NS binaries in an axisymmetric setup. As we show further in Section~\ref{sec:ResMain}, this comparison helps us gain insight into why WD-NS merger simulations with massive WDs have been challenging to perform with $2{\rm D}$ codes.

%\review[inline]{FLASH code short summary (Alexey)}
FLASH is an adaptive mesh refinement (AMR) code that solves the hydrodynamic Euler equations using an unsplit piecewise-parabolic method (${\rm PPM}$) solver. We use ${\rm 2D}$ axisymmetric cylindrical coordinates $[{\bar\rho},\phi,z]$ on a grid of size $(10^{10}$ cm$)\times (10^{10}$ cm), comparable to the orbital size of the WD-NS/BH binaries prior to the merger.

%\review[inline]{Physics components in FLASH (Alexey)}
In our FLASH simulations, we have included source terms for gravity, shear viscosity and the centrifugal force. Self-gravity is included as a multipole expansion of up to $l_{\max}=12-22$ terms using the new FLASH multipole solver. We have added a point-mass gravitational potential to the solver to account for the gravity of the central object. Since our simulations do not include magnetic fields, we cannot self-consistently account for angular momentum transport due to the magneto-rotational instability (MRI). Furthermore, due to the axisymmetric nature of the simulations, the code also cannot capture non-axisymmetric instabilities (e.g. associated with self-gravity) or represent the actual disruption of a three-dimensional white dwarf. To model these processes and to enable the WD disruption in FLASH in the first place, we are using the $\alpha$-viscosity with the parameterization of \citet{Shakura_Sunyaev73}. In this prescription, the kinematic viscosity is:
\begin{equation} 
    \nu_{\alpha}=\alpha C_{s}^{2}/\Omega_{\rm K},  \label{eq:nualpha}
\end{equation}
where $\Omega_{\rm K} = (GM_{\rm enc}/r^{3})^{1/2}$ is the Keplerian frequency given the enclosed mass $M_{\rm enc}$ and $C_{s}$ is the sound speed. In our models, we take $\alpha = 0.1$ for the dimensionless viscosity coefficient in order to facilitate the WD disruption. It is worth noting that, apart from numerical viscosity, dedicated viscosity prescriptions are also available in SPH, e.g. \citet{Rosswog2009,Rosswog2020}. However, we are not using them explicitly in this study because initialising WD disruptions in the 3D case does not require viscosity. As we describe further, in the post-merger evolution of the SPH models, the numerical SPH viscosity effectively represents alpha-viscosity with $\alpha \approx 0.01$. Similarly to the SPH approach, we employ the Helmholz equation of state \citep{Timmes2000}. Since nuclear burning has proven problematic in ${\rm 2D}$ simulations of WD-NS mergers with a massive WD, e.g. \citet{Fernandez19}, we do not include nuclear burning in these simulations. However, we do include neutrino cooling \citep{Chevalier1989,Houck1991} in the internal energy evolution, although the latter does not play an appreciable role for the range of disk radii that we simulate.

%\review[inline]{Model setup in FLASH (Alexey)}
We set up the models closely following \citet{Zenati19a, Zenati19b}. In the setup, similar to the approach used for modelling thermonuclear supernovae \citep{Meakin09}, we initialised the WD as a torus, self-consistently relaxing it with the iterative method introduced in \citet{Zenati19a}. The viscous spreading from the $\alpha$-viscosity term subsequently produces a disc intended to qualitatively resemble the disc from the tidally shredded WD in actual WD-NS/BH mergers. Here we again note that the realistic discs resulting from WD-NS/BH mergers, while being qualitatively similar and having comparable disk heights, differ from the more idealised ones produced in ${\rm 2D}$ simulations \citep{Zenati19b}. A resolution of $8-20$ km was sufficient to achieve $\lesssim$ 10$\%$ convergence in energy for these simulations. To compare the results to those from the SPH code, we calculated the temperature and density distributions in the disc at different times.

\subsection{Nuclear evolution modelling}

%\review[inline]{Nuclear evolution summary (Alexey)}
We model the nuclear evolution for the SPH runs through a nucleosynthetic post-processing step with a large network applied to all the SPH particles we used in the hydrodynamic simulations. This approach is justified because WD-NS binaries do not lead to full detonations \citep{Zenati19a}. Furthermore, local nuclear energy budget in our models is at least several times smaller than the gravitational or kinetic energy budget, as was also the case in \citet{Zenati19b}. The last fact may be understood intuitively because, in WD-NS/BH mergers, nuclear burning happens only relatively close to the NS/BH, deep in its potential well.

%\review[inline]{Nuclear evolution details (Alexey)}
We post-processed the detailed histories of density and temperature from all the SPH particles with a one-zone nuclear Torch code \citep{Timmes1999, Timmes2000b}. We employed a 127-isotope network that includes neutrons and composite reactions from JINA REACLIB \citep{Cyburt2010}. We chose the Torch code over the one-zone MESA burner \citep{Paxton2015} which we used in \citet{Zenati19b} because the latter performed much more slowly in the nuclear statistical equilibrium regime due to a less efficient solver. In the nuclear post-processing, we made sure that the nuclear reactions have not yet started at the beginning of simulations, when the WD is still intact, and have finished by the end of the simulations. This way, we expect our modelling to capture most phases relevant to nucleosynthesis in WD-NS mergers. We have also tested that the final nuclear yields converge in the SPH particle number by re-simulating model W at 20K and 50K resolution. The nuclear code outputs allowed us to calculate the nuclear yields produced in the mergers and were also used to initialise the supernova spectral synthesis simulations.

\subsection{Supernova spectral synthesis}

\label{sec:MethodSNuSetup}

%\review[inline]{SuperNu code introduction (Alexey)}
We map the physical and nuclear compositional properties of WD-NS systems at the end of the simulations as inputs for the openly available radiative transfer code SuperNu \citep{Wollaeger2013,Wollaeger2014}. The SuperNu code stochastically solves the special-relativistic radiative transfer and diffusion equations to first order in $v/c$ in three dimensions by using computationally-efficient hybrid implicit Monte Carlo (IMC) and discrete diffusion Monte Carlo (DDMC) schemes. In this study, we use the spherically-symmetric setup of the SuperNu code to calculate the possible light-curves and spectra expected from WD-NS/BH mergers, similar to how it was done in \citet{Zenati19b}.

%\review[inline]{Mapping to SuperNu, uncertainties  (Alexey)}
SuperNu code models the homologous free expansion phase of supernovae using a velocity grid. The exact mechanism through which the material from the disrupted WD gets ejected into free expansion is still uncertain. In particular, as we discuss in the results Section~\ref{sec:ResHydro}, our 3-dimensional SPH simulations show that the absolute majority of the WD material at the end of the merger is gravitationally bound to the NS/BH and only a small fraction of mass, of order of a fraction of a per cent, gets lost dynamically as tidal ejecta. On the other hand, the gravitational energy budget from the accretion of only a few $0.01\,{\rm M}_\odot$ of material directly onto the surface of the accretor is sufficient to unbind all the remaining material through some feedback mechanism, so long as the energy gets deposited evenly. However, the system is highly asymmetric, which makes the exact fraction of the unbound material uncertain. Therefore, in general, one may expect that some significant fraction of WD material may get ejected after the merger.

%\review[inline]{Mapping to SuperNu, considered scenarios (Alexey)}
Given the uncertainty of how exactly the WD material gets ejected, we map the $3{\rm D}$ velocities and compositions of the SPH particles onto the SuperNu grid following several plausible prescriptions.
1) {\sl Fast ejecta}:~ In this scenario, we map the particle velocities and compositions onto the SuperNu velocity grid as they are, at the end of the simulations. Such a mapping qualitatively represents the case of fast ejecta facilitated by efficient feedback. 2) {\sl Fast mixed ejecta}:~ In this case, we map the particle velocities in the same way as in the previous case. However, each velocity bin in the SuperNu grid now acquires the same mass-averaged composition, as opposed to the previous case when each velocity bin acquired a unique composition based on the SPH particles assigned to that bin. This setup allows us to probe the sensitivity of the results from the previous scenario to the exact velocity distribution of chemical elements. 3) {\sl Slow ejecta}:~ In this scenario, we initialise all the material assuming it was released as a slow wind at low sub-orbital velocities. Specifically, we map the material randomly assigning the element velocities typical values of $400\,{\rm km}/{\rm s}$ with a $1$-$\sigma$ Gaussian spread of $200\,{\rm km}/{\rm s}$. This scenario qualitatively describes the ejection of material through a thermal or mechanical wind, or some generic slow ejection process, with the exact ejection velocities chosen in agreement with \citet{Zenati19b}. If we initialise the Slow ejecta model with additional material dynamically ejected in our SPH simulations, we observe no change to the lightcurve or spectra. The small effect of the dynamical ejecta may be understood by its very low mass, few times $10^{-3}\,{\rm M}_\odot$ and absence of $^{56}$Ni in it. 4) {\sl Reduced-mass ejecta}:~ Here, we initialise the models similar to the Fast mixed ejecta model but reduce the mass in each velocity bin so that the total ejecta mass is $0.7\,{\rm M}_\odot$. Such a setup represents the situation when the feedback only ejects a fraction of the mass efficiently.

%\review[inline]{Other superNu settings (Alexey)}
For each scenario, we set the masses of the SuperNu grid cells equal to the sum of the masses of the corresponding SPH particles. Similarly, we set the chemical compositions and electron fractions $Y_e$ as the mass-weighted average compositions in each cell. We initialised the SuperNu simulations at $0.5\,{\rm d}$ after the merger. By this time, the homologously expanding material adiabatically cools down by many orders of magnitude in temperature. Therefore, we set the temperature of the material in SuperNu initialised to an upper estimate value of $10^4\,\textrm{K}$, any lower values essentially leading to the same lightcurves or spectra. We construct the photometric lightcurves in the AB magnitude system from spectral energy distributions (SEDs) by using LSST ugrizy filters provided by virtual observatory SED analyser tool VOSA \citep{Bayo2008}.

\section{Results}
\label{sec:ResMain}

%\review[inline]{Short review of the results (Alexey)}
In this section, we present the results from our modelling of ONe WD-NS/BH mergers. In Section~\ref{sec:ResHydro}, we describe the hydrodynamic evolution, as seen from our $3{\rm D}$ SPH simulations. Then, in Section~\ref{sec:ResNuc}, we describe the nuclear evolution, as follows from post-processing of our SPH simulations with nuclear Torch code. We then present the yields of nuclear elements resulting from ONe WD-NS/BH mergers and their dependence on the WD mass. In Section~\ref{sec:ResTransients}, we use these nuclear yields together with the SuperNu code to construct the lightcurves and spectra of the transient events resulting from the mergers. Since it is still unknown how exactly the material gets ejected from merging WD-NS binaries, we consider several most likely scenarios and outline several likely resulting transients. Finally, in Section~\ref{sec:Res2D3DComparison}, we compare our hydrodynamic SPH simulations to $2{\rm D}$ non-nuclear FLASH code simulations of the same mergers.

\subsection{Hydrodynamic evolution}

\label{sec:ResHydro}

\begin{figure*}
\includegraphics[width=0.8\linewidth]{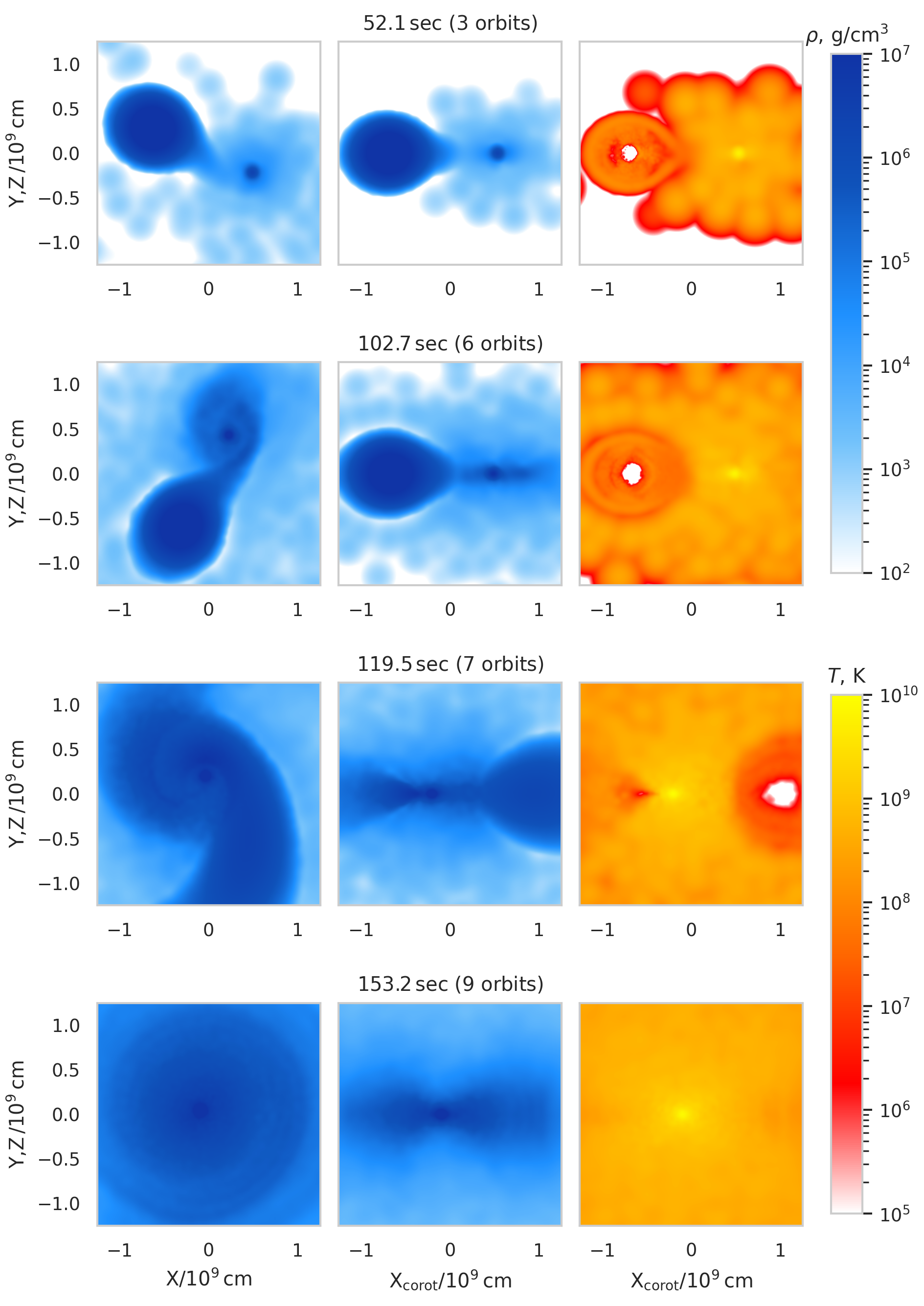}
\caption{Snapshots from the SPH simulation of model W, which represents a typical ONe WD-NS binary with a $1.1\,{\rm M}_\odot$ ONe WD and a $1.4\,{\rm M}_\odot$ NS. The three columns show the density in the horizontal ${\rm X}$--${\rm Y}$ plane and the density and temperature in the vertical corotating ${\rm X}_{\rm corot}$--${\rm Z}$ plane, respectively. The four rows show the binary at different stages of the merger. By the start of the simulation, the donor has been relaxed into a state when the mass-transfer stream has already developed. The snapshots in the first row, at about $52.1\,{\rm sec}$ (3 initial orbital periods) into the simulation, show the mass transferring WD that is still intact. At this point, the mass transfer rate grows steadily. The snapshots in the second row show the deformed WD shortly before its disruption, at $102.7\,{\rm sec}$ (6 initial periods) into the simulation. Only minor nuclear synthesis occurs until this point. The snapshots in the third row, taken at about $119.5\,{\rm sec}$ (7 initial periods), show the WD in the process of being tidally shredded. The self-crossings of the tidally stretched WD produce shocks, heating the material and initiating significant nuclear burning. The snapshots in the fourth row show the WD shortly after the disruption, at about $153.2\,{\rm sec}$ (9 initial periods) into the simulation. By this time, the WD material has formed a disc surrounded by a tenuous and hot corona, the bulk of the nuclear material has been synthesised. In the subsequent ten orbits of the simulation, nuclear burning subsides, and the disc structure remains nearly unchanged.}
\label{fig:SPH}
\end{figure*}

\begin{figure*}
\includegraphics[width=\linewidth]{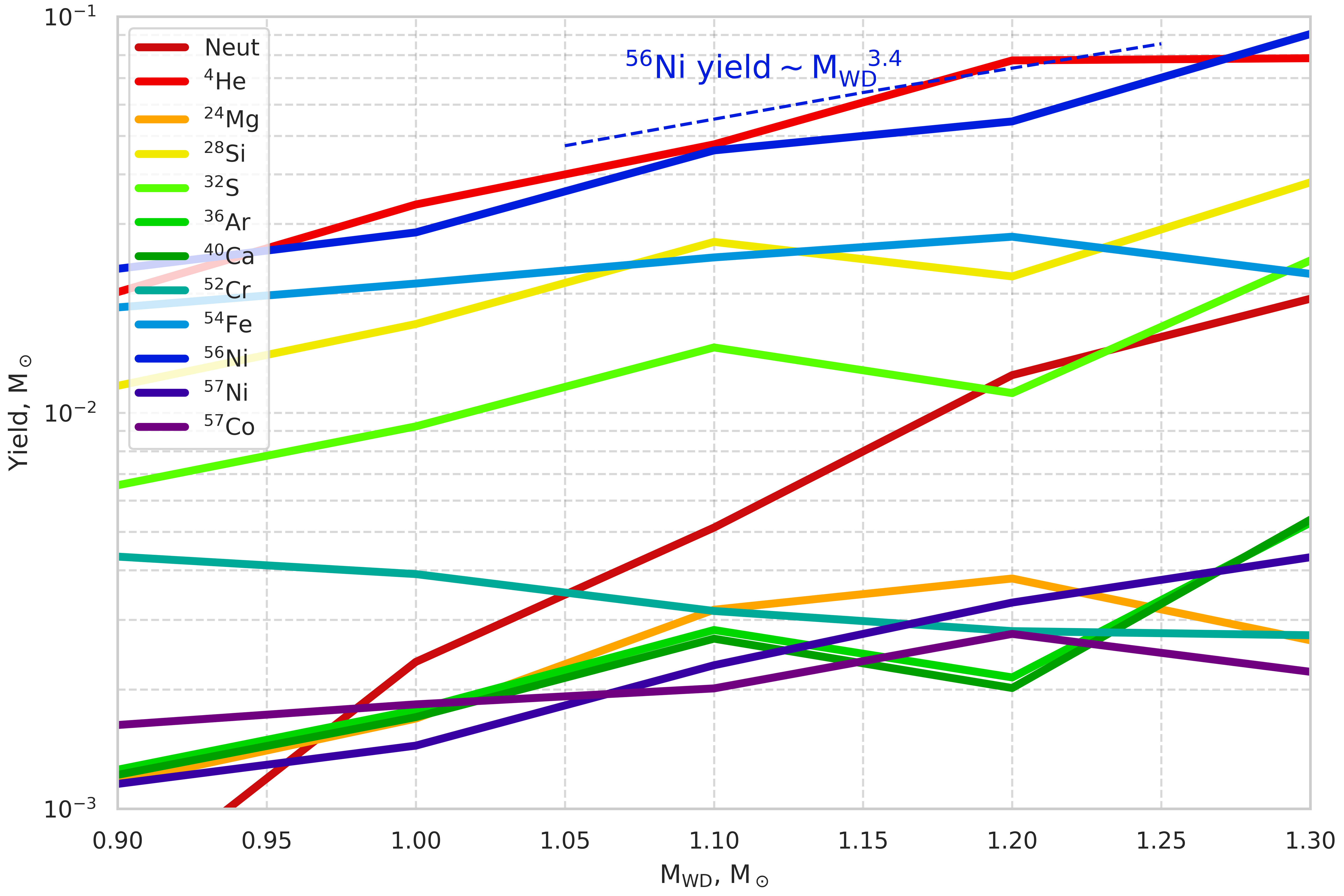}
\caption{The nuclear yields synthesised at the end of our SPH simulations as a function of mass (based on models U, V, W, X, Y). Alpha elements up till and including $^{40}$Ca, as well as $^{56}$Ni and $^{57}$Ni show a strong $\sim$M$_{\rm WD}^{~~~~3.4}$ sensitivity to the WD mass (the power-law fit is for $^{56}$Ni). As a result, ONe WD-NS mergers likely produce relatively bright transients and dominate the pollution of the interstellar environment compared to all WD-NS binaries. Some other elements, for example, $^{52}$Cr, $^{54}$Fe, $^{57}$Co show a weaker sensitivity to the WD mass.}
\label{fig:SPHYields}
\end{figure*}

\begin{table*}
\begin{tabular}{|c|c|c|c|c|c|c|c|c|}
\hline
          Property & Model U & Model U$_2$ & Model V & Model W & Model X & Model X$_2$ & Model X$_3$ & Model Y \\
\hline
 M$_{\rm WD}$, M$_\odot$ &  0.9000 &   0.9000 &  1.0000 &  1.1000 &  1.2000 &   1.2000 &   1.2000 &  1.3000 \\
    M$_{\rm comp}$, M$_\odot$ &  1.4000 &   1.4000 &  1.4000 &  1.4000 &  1.4000 &   2.0000 &   5.0000 &  1.4000 \\
\hline
         $^{}$Neutrons &  0.0006 &   0.0006 &  0.0023 &  0.0051 &  0.0124 &   0.0029 &   0.0009 &  0.0194 \\
         $^1$H &  0.0003 &   0.0003 &  0.0007 &  0.0010 &  0.0015 &   0.0011 &   0.0014 &  0.0015 \\
          $^{4}$He &  0.0202 &   0.0202 &  0.0336 &  0.0476 &  0.0775 &   0.0285 &   0.0091 &  0.0785 \\
          $^{12}$C &  0.0000 &   0.3540 &  0.0000 &  0.0000 &  0.0000 &   0.0000 &   0.0000 &  0.0000 \\
          $^{16}$O &  0.4553 &   0.3724 &  0.4927 &  0.5168 &  0.5424 &   0.6662 &   0.7069 &  0.6690 \\
         $^{20}$Ne &  0.2960 &   0.0040 &  0.3171 &  0.3241 &  0.3386 &   0.4183 &   0.4589 &  0.2090 \\
         $^{24}$Mg &  0.0012 &   0.0060 &  0.0017 &  0.0032 &  0.0038 &   0.0048 &   0.0026 &  0.0027 \\
         $^{28}$Si &  0.0117 &   0.0227 &  0.0167 &  0.0270 &  0.0221 &   0.0147 &   0.0047 &  0.0382 \\
          $^{32}$S &  0.0066 &   0.0074 &  0.0092 &  0.0146 &  0.0112 &   0.0063 &   0.0017 &  0.0242 \\
         $^{36}$Ar &  0.0013 &   0.0011 &  0.0018 &  0.0028 &  0.0021 &   0.0012 &   0.0004 &  0.0053 \\
         $^{40}$Ca &  0.0012 &   0.0009 &  0.0017 &  0.0027 &  0.0020 &   0.0012 &   0.0003 &  0.0054 \\
         $^{49}$Ti &  0.0002 &   0.0002 &  0.0002 &  0.0002 &  0.0018 &   0.0000 &   0.0000 &  0.0070 \\
          $^{51}$V &  0.0012 &   0.0012 &  0.0011 &  0.0009 &  0.0040 &   0.0001 &   0.0000 &  0.0141 \\
         $^{52}$Cr &  0.0043 &   0.0043 &  0.0039 &  0.0032 &  0.0028 &   0.0003 &   0.0000 &  0.0027 \\
         $^{53}$Mn &  0.0014 &   0.0014 &  0.0014 &  0.0014 &  0.0019 &   0.0003 &   0.0000 &  0.0015 \\
         $^{54}$Mn &  0.0013 &   0.0013 &  0.0012 &  0.0011 &  0.0011 &   0.0001 &   0.0000 &  0.0012 \\
         $^{55}$Mn &  0.0030 &   0.0030 &  0.0026 &  0.0021 &  0.0038 &   0.0002 &   0.0000 &  0.0134 \\
         $^{54}$Fe &  0.0184 &   0.0188 &  0.0212 &  0.0247 &  0.0278 &   0.0022 &   0.0005 &  0.0224 \\
         $^{55}$Fe &  0.0029 &   0.0029 &  0.0032 &  0.0033 &  0.0041 &   0.0005 &   0.0001 &  0.0031 \\
         $^{56}$Fe &  0.0239 &   0.0239 &  0.0244 &  0.0215 &  0.0157 &   0.0008 &   0.0001 &  0.0140 \\
         $^{56}$Ni &  0.0231 &   0.0264 &  0.0285 &  0.0460 &  0.0544 &   0.0340 &   0.0101 &  0.0904 \\
         $^{57}$Ni &  0.0012 &   0.0013 &  0.0014 &  0.0023 &  0.0033 &   0.0020 &   0.0004 &  0.0043 \\
         $^{58}$Ni &  0.0149 &   0.0157 &  0.0218 &  0.0359 &  0.0510 &   0.0116 &   0.0013 &  0.0547 \\
         $^{57}$Co &  0.0016 &   0.0016 &  0.0018 &  0.0020 &  0.0028 &   0.0003 &   0.0000 &  0.0022 \\
\hline
\end{tabular}
\caption{The table presents the nuclear elements, in solar masses, at the end of each simulation. The largest yields are, typically, produced for alpha-elements and near the iron peak, whereas for the other elements, the production yields are below $10^{-4}\,$M$_\odot$. One may obtain characteristic Galactic enrichment rates from WD-NS binaries by multiplying the yields for model W (the typical ONe WD-NS binary) by $200\,{\rm Myr}^{-1}$, the characteristic inspiral rate for the population.}
\label{tab:Nucyields}
\end{table*}

%\review[inline]{Initial conditions are realistic (Alexey)}
We present example snapshots from our SPH simulation for a $1.1 + 1.4\,{\rm M}_\odot$ ONe WD-NS binary (model W) in Figure~\ref{fig:SPH}. At the beginning of the simulations, we initialised the binaries when the WD is still intact. As we describe in Section~\ref{sec:MethodSPHSetup}, the WD has been relaxed in a binary with the mass transfer stream already open, so that no spurious WD oscillations occur at the start of the main simulations. The mass transfer rate, from this point, continually grows until the WD gets disrupted at about $5-10$ orbital periods into the simulation. The beginning of the simulations matches the late stages of evolution of real WD-NS binaries which start transferring mass at yet lower rates. Therefore, we expect that by the time of the merger, the WD model, the mass transfer stream and the material around the neutron star accurately represent real physical mergers.

%\review[inline]{Disruption morphology (Alexey)}
The disruption of the WD proceeds similarly as in our earlier $3{\rm D}$ modelling of lower-mass CO WD-NS mergers \citep{Zenati19b}. The actual tidal shredding takes place over about one orbital period, and the WD material circularises over an additional period. The circularisation happens because the shocks, caused by self-intersections of the stream, preserve the orbital angular momentum and turn the orbital energy into heat. The disc from the WD remnant, for that reason, is puffed up and has a thickness comparable to the radius of the original WD, being more concentrated near the midplane. The disc material is shock-heated the most within the circularisation radius, and even more so around the NS. A $6\times6\times4$ binary orbits-sized sparse corona of hot material, making about $20$ per cent of the WD mass, surrounds the remnant object. However, the amount of material becoming gravitationally unbound and dynamically ejected during the merger is very small, a fraction of a per cent of the WD material. This material is ejected with velocities of $10$--$20$ thousands of ${\rm km}/{\rm s}$. The disc changes very little during the ten orbital periods after the merger, getting slightly hotter and denser in the innermost regions and slightly spreading radially. By considering the evolution of the average radius and the velocity dispersion of the disc, we estimate its viscous timescale to be approximately $50$ orbital periods, which corresponds to an effective alpha-viscosity parameter of $0.008$ and is mostly driven by numerical viscosity. This effective alpha-viscosity mimics the qualitative effects of the MRI-viscosity likely produced following the WD-NS/BH disruptions in real binaries. However, three-dimensional magnetohydrodynamic simulations would be needed to model MRI viscosity in full detail.

%\review[inline]{Trends in the hydrodynamic models (Alexey)}
Our WD-NS models containing more massive WDs generally have shorter orbital periods and larger binding energies. The trend arises because more massive WDs are more compact and can fit in a more compact orbit, in addition to the direct gravitational effect of the WD mass itself \citep{Metzger2012}. As a result, mergers with more massive WDs release more heat energy in a shorter time and a smaller volume, leading to higher disk densities and temperatures, as observed in our simulations.

\subsection{Nuclear evolution}
\label{sec:ResNuc}

%\review[inline]{Nuclear burning process, general (Alexey)}
In terms of nuclear evolution, models with massive ONe WDs may be seen as more energetic but otherwise similar versions of CO WD-NS mergers. Nuclear burning sets in during the phases when the WD is still intact and fully subsides by the end of our simulations. The bulk of the new elements gets synthesised during the early disc phase, which lasts about two orbital periods. This active burning phase starts when the WD is shredded into a long self-crossing filament and ends roughly when the initially eccentric disc material settles into a steady circular state. Nuclear synthesis is happening inside the circularisation radius. For example, in model W, nuclear burning happens at distances of the order of $700\,{\rm km}$ from the NS accretor. At the same time, the characteristic size of SPH particles in this region, given by their kernel core radius, is of the order of $70\,{\rm km}$. Thus, the nuclear burning region is resolved well. The fact that nuclear burning is resolved may also be seen since nuclear burning affects ten to twenty thousand SPH particles depending on the model, or about ten to twenty per cent of the WD mass. The production of $^{56}{\rm Ni}$ persists for several periods longer into the disc evolution at a slower rate, in the vicinity of the NS, and eventually fully subsides. The innermost regions of the disc reach sufficiently high temperatures and densities that photo-disintegration takes place, producing about $0.02\,{\rm M}_\odot$ to about $0.1\,{\rm M}_\odot$ of $^4{\rm He}$, as well as between $7\cdot10^{-4}\,{\rm M}_\odot$ and $0.02\,{\rm M}_\odot$ of free neutrons, in strong dependence on the WD mass. The free neutrons may get captured by other elements shortly after the merger, thus modifying up to one per cent of all burning products, or decay into hydrogen. However, due to the low total mass of free neutrons, the r-process, in which the seed $^{56}{\rm Ni}$ nuclei rapidly capture significant numbers of free neutrons, is very unlikely to happen. Nuclear reactions happen deeply in the potential well of the NS, have a small energy budget compared to the gravitational binding energy of the particles and are therefore dynamically unimportant. The low dynamical importance of nuclear reactions, in this case, justifies our approach of applying nuclear evolution only in the post-processing stage.

%\review[inline]{Nuclear yields are strongly sensitive to WD mass (Alexey)}
Overall, ONe WD-NS mergers produce almost an order of magnitude larger amounts of most elements than CO WD-NS mergers. We summarise the trend in nuclear yields with mass immediately after the merger for all our models in Figure~\ref{fig:SPHYields}. The yield for $^{56}{\rm Ni}$ depends on the donor mass as $\sim$~M$_{\rm WD}^{~~~~3.4}$, based on a power-law fit. Other alpha-elements up to $^{40}{\rm Ca}$ as well as $^4{\rm He}$ show a similar trend. The scaling is stronger than E$_{\rm kin}/P_{\rm orb} \sim $~M$_{\rm WD}^{~~~~2}$, the typical rate of conversion of kinetic energy into heat during the merger, and is likely related to the sensitivity of nuclear reaction rates to the temperature and density. Additionally, we observe that more massive WDs lead to more abrupt production of the bulk of the elements.

%\review[inline]{Production of interesting nuclear elements (Alexey)}
ONe WD-NS mergers dominate the Galactic nuclear element production from WD-NS binaries for most elements, due to their large nuclear yields and the inspiral rates comparable to other WD-NS binaries as discussed in Section~\ref{sec:Intro}. We provide a detailed summary for all the models, and an extended set of elements, in Table~\ref{tab:Nucyields}. From the table, we see that these mergers typically produce a few times $0.01\,{\rm M}_\odot$ of $^{28}{\rm Si}$ and $^{32}{\rm S}$ and a few times $0.001\,{\rm M}_\odot$ of $^{36}{\rm Ar}$, $^{40}{\rm Ca}$ and $^{52}{\rm Cr}$. The models also produce between $0.02\,{\rm M}_\odot$ and $0.03\,{\rm M}_\odot$ of $^{54}{\rm Fe}$ and a comparable amount of $^{56}{\rm Fe}$, which have been used as a diagnostic for proposing the ONe WD-NS nature for SN AT2018kzr by \citet{Gillanders2020}. We also find that the models produce between $0.001\,{\rm M}_\odot$ and $0.003\,{\rm M}_\odot$ of $^{57}$Ni and $^{57}$Co, which have been used to constrain the progenitor nature of Ca-rich transient SN 2019ehk from its late-time photometry \citet{JacobsonGalan2020, JacobsonGalan2020b}. $^{44}{\rm Ti}$, which may lead to the diffuse Galactic positron background, e.g. \citet{Crocker2017}, is only produced in small amounts, up to a few times $10^{-6}\,{\rm M}_\odot$. One may use Table~\ref{tab:Nucyields} to obtain the characteristic Galaxy enrichment rates by WD-NS binaries. In this case, one should multiply the rate from model W, which represents the typical ONe WD-NS binary, by the empirical WD-NS inspiral rate of $200\,{\rm Myr}^{-1}$  per Milky Way-like galaxy \citep{Toonen2018, Bobrick2017}.

%\review[inline]{Sensitivity of nuclear yields on the accretor mass and the initial WD composition (Alexey)}
By comparing the nuclear yields for models U and U$_2$, which only differ by the initial composition, we see that the initial composition has a minor effect on all the produced elements. The exception is $^{24}$Mg and $^{28}$Si, where the yields differ by a factor of two. Therefore, as expected, the initial ONe composition by itself likely does not make the simulations of such binaries more challenging. By comparing the yields for models X, X$_2$ and X$_3$, which differ only by the accretor mass (1.4, 2.0 and 5.0 M$_\odot$, correspondingly), we see that the nuclear production of $^{56}$Ni decreases almost in inverse proportion with the accretor mass and also generally decreases for alpha-elements. The less vigorous nuclear element production likely explains why ONe WD binaries with a massive BH companion, such as the ones modelled by \citet{Fernandez19}, are less challenging to simulate compared to the binaries with lower-mass NS companions.

%\review[inline]{Cautionary note on using the numbers}
The nuclear yields summarised in Table~\ref{tab:Nucyields} describe the elements produced at the end of the merger simulations. In all currently existing models of WD-NS/BH mergers (in this study and the literature), it is not entirely certain what fraction of the WD material eventually gets ejected from the system. While we argue in Section~\ref{sec:Disc} that likely most of the WD material is indeed ejected, the abundances in Figure~\ref{fig:SPHYields} and Table~\ref{tab:Nucyields} should be considered only as upper limits on the amount of the actually ejected elements.

\begin{figure}
\includegraphics[width=\linewidth]{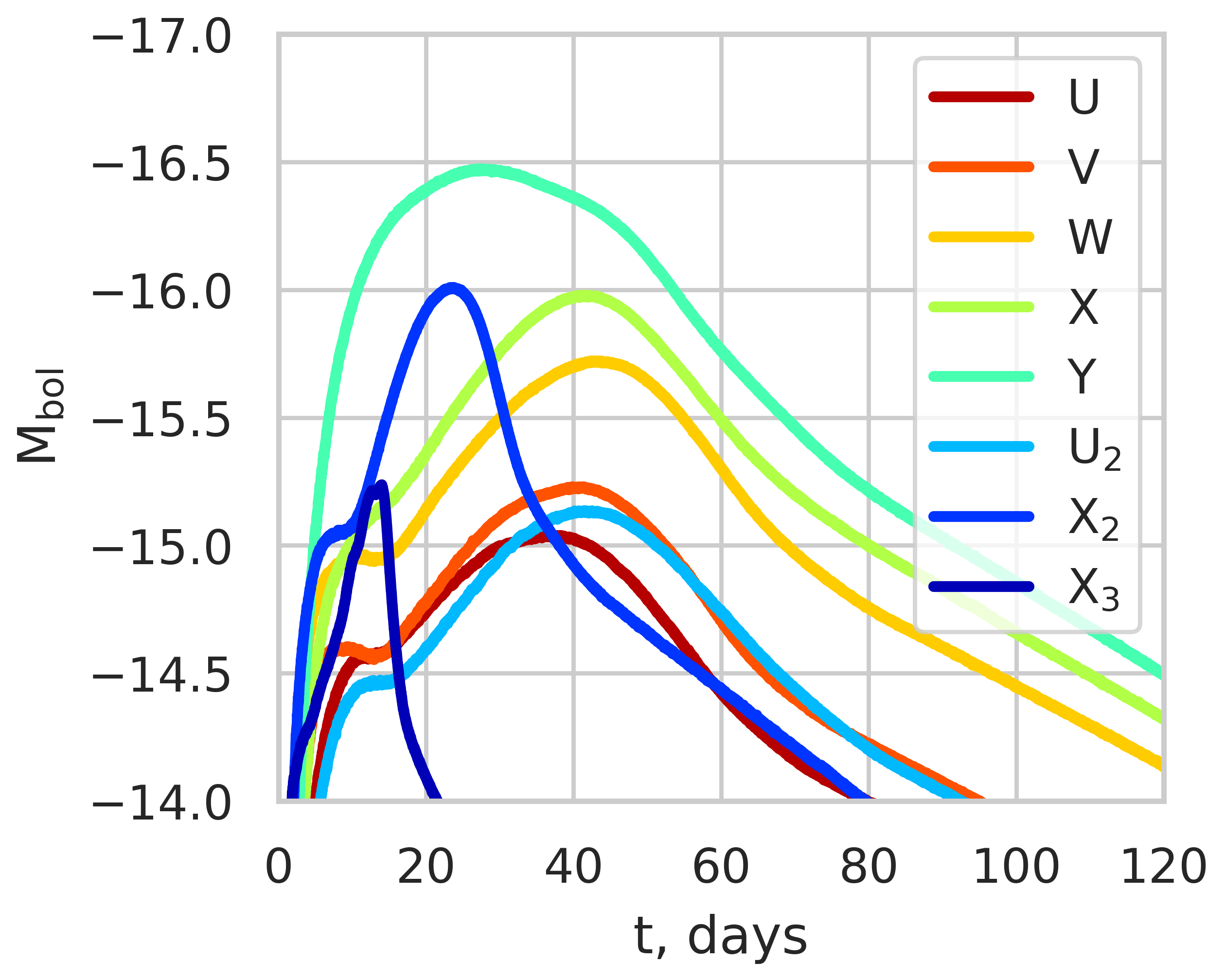}
\caption{Bolometric lightcurves for the transients resulting from WD-NS/BH mergers, for the fiducial ({\sl Fast Ejecta}) model of mass ejection. Models U, V, W, X and Y show a sequence of ONe WD-NS mergers with a $1.4\,$M$_\odot$ NS and the WD mass increasing in equal steps from $0.9$ to $1.3\,$M$_\odot$. Model U$_2$ shows the effect of replacing the WD composition by CO in model U, and models X$_2$ and X$_3$ show the effect of replacing the accretor with a more massive $2.0\,$M$_\odot$ NS and $5.0\,$M$_\odot$ BH compared to model X, respectively. One may see that for higher-mass WDs, the resulting transients are generally longer and brighter.}
\label{fig:LCFiducial}
\end{figure}

\begin{figure*}
\includegraphics[width=0.88\linewidth]{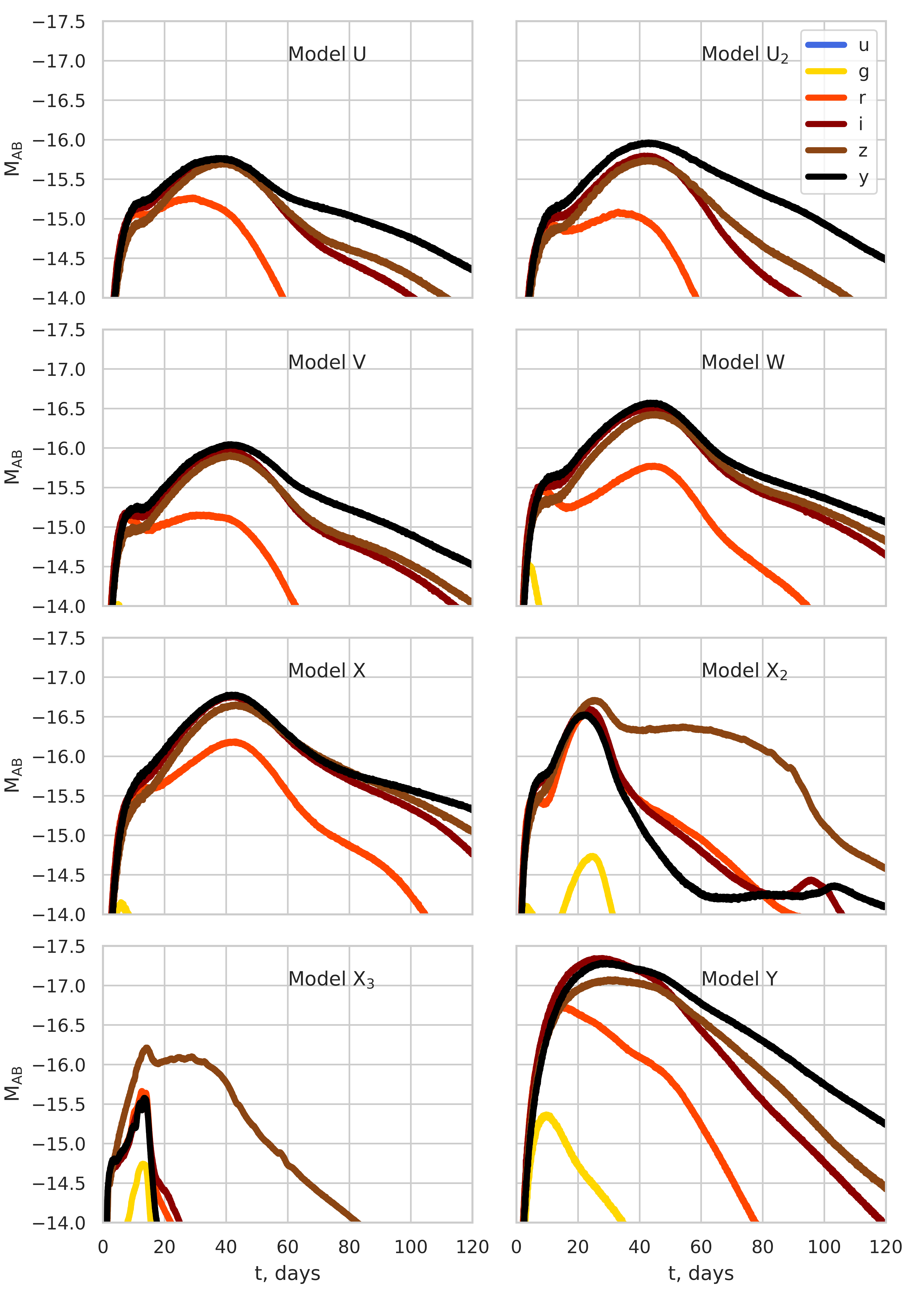}
\caption{Photometric evolution in Vera Rubin observatory's (LSST) {\sl ugrizy} bands, in the AB magnitude system, for the transients resulting from WD-NS/BH mergers, for the fiducial ({\sl Fast Ejecta}) model of mass ejection. Models U, V, W, X and Y show a sequence of ONe WD-NS mergers with a $1.4\,$M$_\odot$ NS and the WD mass increasing in equal steps from $0.9$ to $1.3\,$M$_\odot$. Model U$_2$ shows the effect of replacing the WD composition by CO in model U, and models X$_2$ and X$_3$ show the effect of replacing the accretor with a more massive $2.0\,$M$_\odot$ NS and $5.0\,$M$_\odot$ BH compared to model X, respectively. The plots show that all the transients are dominantly red/infrared.}
\label{fig:LCFiducial2}
\end{figure*}

\begin{figure*}
\includegraphics[width=0.88\linewidth]{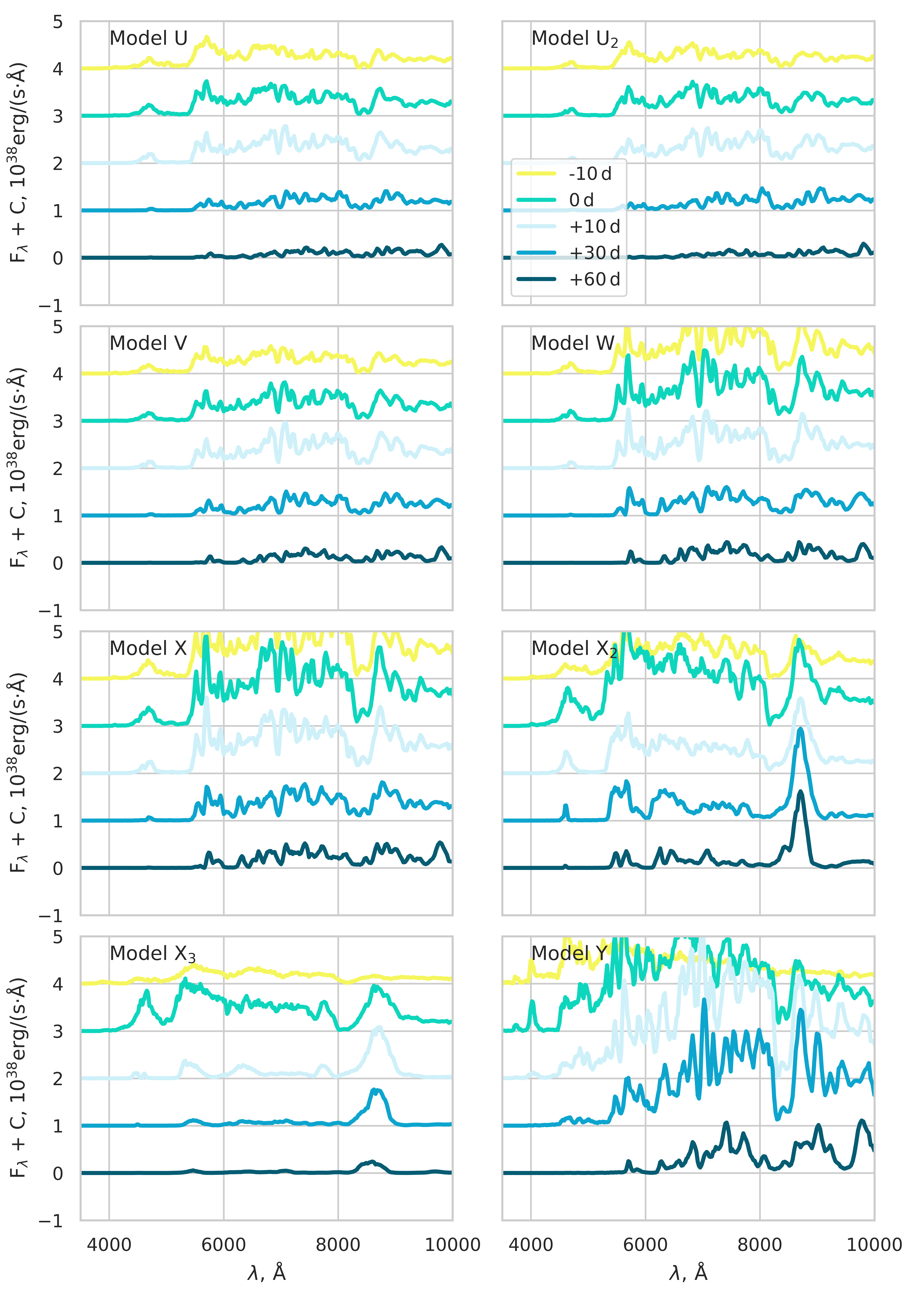}
\caption{Spectral densities for the transients resulting from WD-NS/BH mergers, for the fiducial ({\sl Fast Ejecta}) model of the mass ejection, taken at $-10$, $0$, $10$, $30$ and $60$ days relative to the r-band peak. Models U, V, W, X and Y contain a $1.4\,$M$_\odot$ NS and the ONe WD mass increasing in equal steps from $0.9$ to $1.3\,$M$_\odot$. Model U$_2$ shows the effect of replacing the WD composition by CO in model U, and models X$_2$ and X$_3$ show the effect of replacing the accretor with a more massive $2.0\,$M$_\odot$ NS and $5.0\,$M$_\odot$ BH compared to model X, respectively.}
\label{fig:SpectraFiducial}
\end{figure*}

\begin{figure*}
\includegraphics[width=0.84\linewidth]{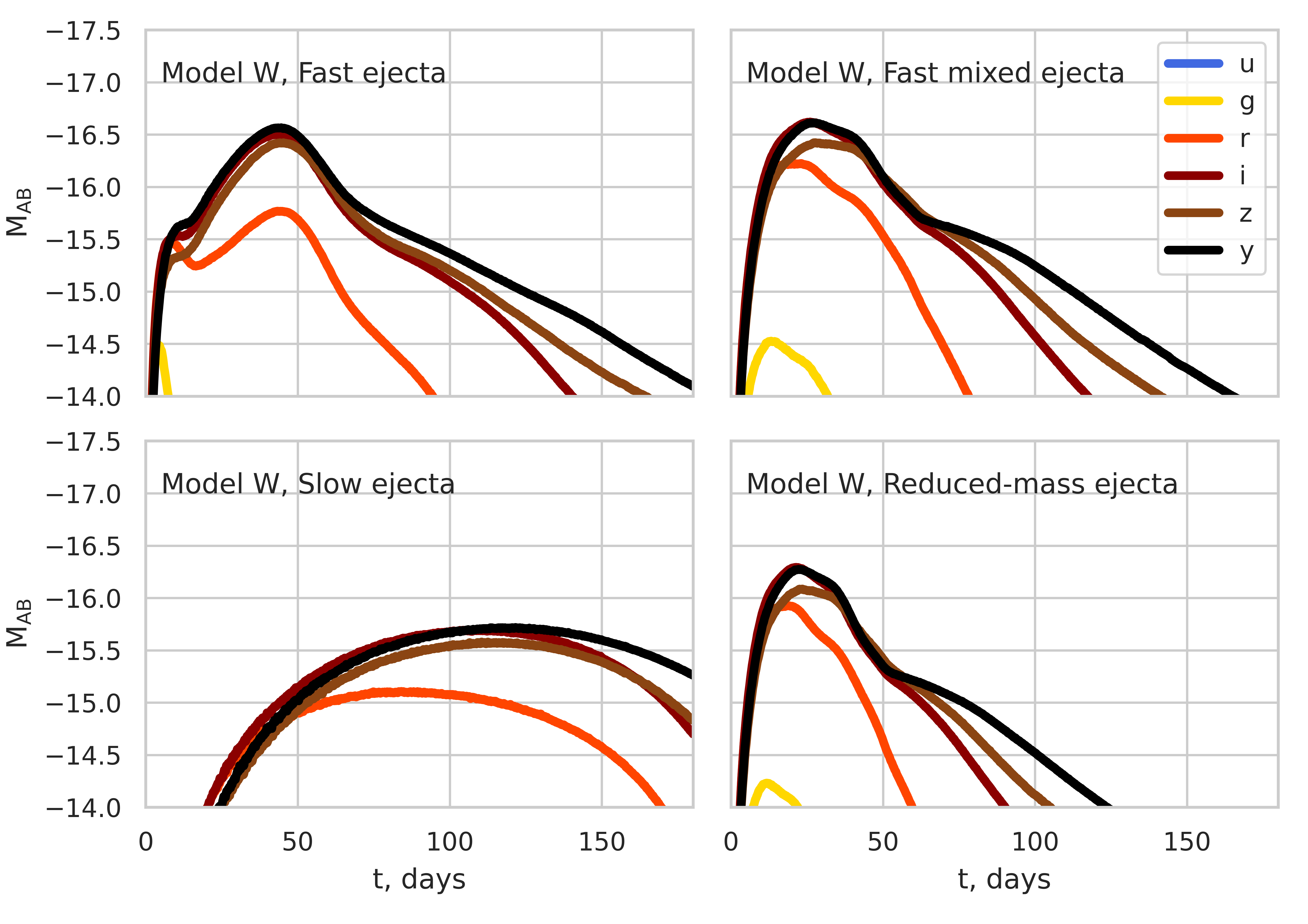}
\includegraphics[width=0.84\linewidth]{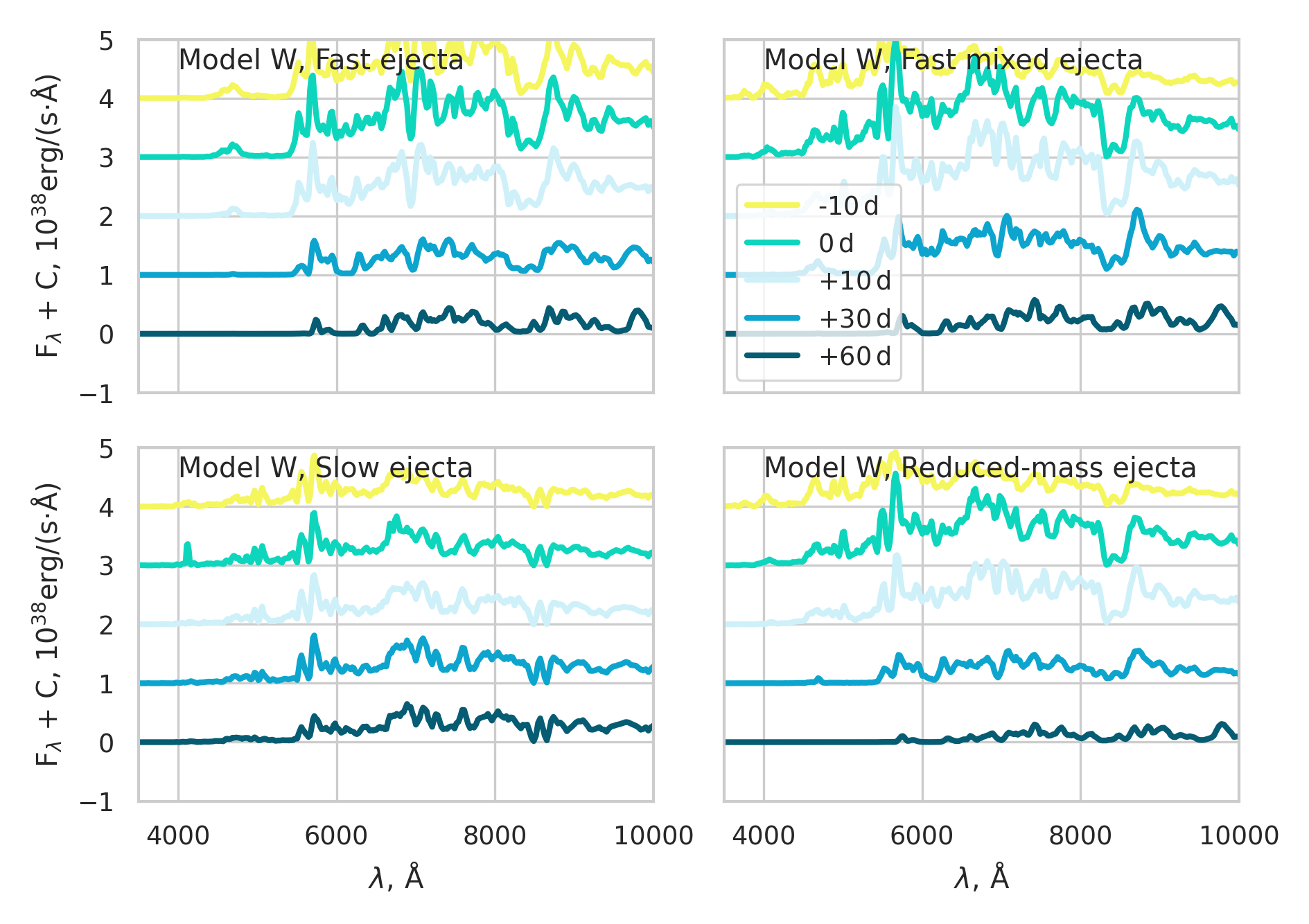}    
\caption{Comparison of the effect of different mass ejection models on the lightcurves (top two rows) and spectra (bottom two rows) from WD-NS/BH mergers. In all the plots, we use model W, corresponding to a typical $1.1\,$M$_\odot$ ONe WD + $1.4\,$M$_\odot$ NS binary. We show examples for the {\sl Fast ejecta} model (material ejected with velocities and compositions as is), {\sl Fast mixed ejecta} model (same as before, but compositions are homogeneously mixed), {\sl Slow ejecta} model (material is ejected as a slow wind) and {\sl Reduced-mass ejecta} model (same as Fast mixed ejecta model, but only $0.7\,{\rm M}_\odot$ is ejected). The figures show that compositional mixing may affect some spectral features, with more mixed models leading to smoother but otherwise similar lightcurves. The slow ejecta model makes the transients slower-evolving and dimmer, with spectra resembling later-time spectra of the faster-evolving model. Reducing the ejecta mass while preserving the composition leads to fainter and faster-evolving transients with similar spectra.
}
\label{fig:ModelComparison}
\end{figure*}

\subsection{Lightcurves and spectra}
~\label{sec:ResTransients}

%\review[inline]{Short summary of the SuperNu modelling (Alexey)}
In this section, we present lightcurves and spectra obtained from our radiative transfer modelling with SuperNu code. At the end of our simulations, most of the material remains gravitationally bound, while the mechanism for mass ejection in ONe WD-NS/BH mergers is presently unknown. Therefore, we consider several most plausible scenarios of mass ejection. This way, we construct the likely range of lightcurves and spectra expected from these mergers.

%\review[inline]{Fiducial model: bolometric lightcurves (Alexey)}
In our fiducial model ({\sl Fast Ejecta} model), we assume that the WD material is ejected with the velocities and compositions as they are at the end of our merger simulations, similarly to our earlier modelling in \citet{Zenati19b}. In other words, the fiducial model assumes that the material gets ejected through an efficient feedback process. In Figure~\ref{fig:LCFiducial}, we show the bolometric lightcurves expected for different models in Table~\ref{tab:Models}. Since the peak luminosity scales linearly with the amount of $^{56}$Ni \citep{Arnett1979}, we observe that the Ni-rich ONe WD-NS model Y with the most massive $1.3\,$M$_\odot$ WD reaches $-16.5$ magnitudes at the peak. The transient is about 1.5 magnitudes brighter than for model U with the least massive $0.9\,$M$_\odot$ WD and about 2.5 magnitudes brighter than in our model D in \citet{Zenati19b} with a $0.62$M$_\odot$ CO WD. Furthermore, since the photon diffusion time scales as $M_{\rm ej}$ \citep{Arnett1979}, we observe that ONe WD-NS mergers result in transients with about two times larger peak widths than CO WD-NS mergers. Faster rise times in the models with more massive WDs may be explained by the higher material ejection velocities, in turn coming from the higher kinetic energy in such models.

%\review[inline]{Fiducial model: photometry (Alexey)}
We show the photometric lightcurves in the Vera Rubin Observatory's (LSST) {\sl ugrizy} bands in the AB magnitude system for all our WD-NS/BH models and the fiducial mass ejection model in Figure~\ref{fig:LCFiducial2}. The evolution is typical to that of nuclear-powered transients. The red and infrared bands dominate the lightcurve near the peak. The colour may be explained by the relatively small amount of nickel-56 and hence relatively low temperature in these transients. In comparison, more nickel-rich type Ia supernovae, which have otherwise comparable ejecta masses and velocities, are white near the peak. The fiducial models show an early bump at few days in the red and visual bands. The bump is because, in our fiducial models, a significant fraction of $^{56}$Ni is present in the fastest moving ejecta. The photon diffusion time from the outer ejecta is low since the photons need to traverse only a small amount of mass to escape. As a result, the radiation from outer $^{56}$Ni escapes early on and when the ejecta has not yet cooled down by expansion. The WD-NS transients with different WD masses (models U, V, W, X, Y) show similar colour evolution, also consistent with that from lower-mass CO WD-NS mergers \citep{Zenati19b}. The model based on CO WD-NS composition (model U$_2$), expectedly, shows a nearly identical lightcurve compared to the ONe model U, slightly modified by the carbon opacity. Model X$_2$ with a $2.0\,$M$_\odot$ NS is almost two times faster-evolved in most bands compared to model X with a $1.4\,$M$_\odot$ NS, due to faster ejecta velocity. Model X$_3$, with a BH-WD binary, produces yet faster ejecta and additionally contains most of its $^{56}$Ni in the fastest ejecta. As a result, the model produces a sharp early red peak, followed by a few weeks-wide infrared decay.

%\review[inline]{Fiducial model: Spectra (Alexey)}
In Figure~\ref{fig:SpectraFiducial}, we show the evolution of spectra from all the WD-NS/BH models for the fiducial model of mass ejection. For all the models, spectral densities correspond to the epochs at $-10$, $0$, $10$, $30$ and $60$ days relative to the r-band maximum. The blue and visual parts of the spectrum for WD-NS mergers (models U, V, W, X, Y) are suppressed, probably because of the continuum and their relatively low nickel content. The red and the IR parts of the spectrum show strong line absorption. The spectra of ONe and CO WD disruptions (models U and U$_2$) are qualitatively similar, with the CO model showing stronger absorption in visual at the early time. Higher-mass models X and Y show weaker absorption near $5000\,{\rm \text{\normalfont\AA}}$ at the early times, which can probably be related to their higher photospheric temperatures. The model with a massive NS (model X$_2$) shows weaker line absorption in red near the peak, while the ONe WD-BH model (model X$_3$) shows strong absorption different from other models.

%\review[inline]{Effects of compositional mixing (Alexey)}
The different models of how mass is ejected in our binaries, described in Section~\ref{sec:MethodSNuSetup}, allow us to assess the uncertainties in our current lightcurve and spectral modelling. We show the typical effects in Figure~\ref{fig:ModelComparison}. The fiducial {\sl Fast Ejecta} model, considered in this Section, is based on the assumption that the material is ejected from the disk while preserving its velocity and composition structure. In {\sl Fast mixed ejecta} model, we initialise the compositions assuming they get homogeneously mixed instead, allowing us to assess the effects of mixing on the lightcurves and spectra. The main effect of mixing is that the lightcurves become smoother. In particular, the early bump in the r-band disappears for all models, which is expected since the bump is caused by excess $^{56}$Ni in the outer ejecta. The peak in the mixed models occurs a bit earlier, likely also related to nickel redistribution. The absorption lines in the mixed models mostly get weaker compared to the fiducial model. This behaviour may be explained by a smaller amount of burned material being outside the photosphere for these models. We observe similar effects for all the other WD-NS/BH models.

%\review[inline]{Effects of the ejection velocities (Alexey)}
The velocity distribution of the ejected material is also uncertain. In the {\sl Slow ejecta} model, we investigate the characteristic effect of the ejecta velocities being small. For this, we adopt material velocities with a Gaussian mean of $400\,$km/s and a 1-$\sigma$ spread of $200\,$km/s, as would occur in mechanically-driven disc winds. In this case, the compositions are effectively initialised as mixed. As we see in Figure~\ref{fig:ModelComparison}, the main effect on the lightcurves is that the transient evolution is several times slower, in proportion to the mean velocity. The transient peaks at about $100$ days, showing only weak dependence on the WD mass. Since the radiated energy must remain the same, the resulting transients are typically about one magnitude dimmer. Due to the same velocity distribution, the lightcurve shapes for all WD-NS/BH models are similar, only showing differences in the peak luminosity, in proportion to the amount of $^{56}$Ni. The spectra of slow-ejecta models partly resemble later-time spectra of faster-evolving models.

Finally, the {\sl Reduced-mass ejecta} model may be seen as a fainter and faster-evolving version of the {\sl Fast ejecta} and {\sl Fast mixed ejecta} models. In the {\sl Reduced-mass ejecta} model, the material is initialised similarly as in the {\sl Fast mixed ejecta} model. However, we decreased the ejecta mass in each velocity bin proportionally while preserving the composition, so that the total ejecta mass is $0.7\,{\rm M}_\odot$ for model W, rather than $1.1\,{\rm M}_\odot$. The peak luminosity becomes smaller in proportion to the mass due to a smaller amount of $^{56}$Ni. The evolution becomes faster in proportion to $M_{\rm ej}$, as expected. Due to the same composition and velocity structure, however, spectral evolution is very similar to the fast ejecta models.

\begin{figure*}
\includegraphics[width=\linewidth]{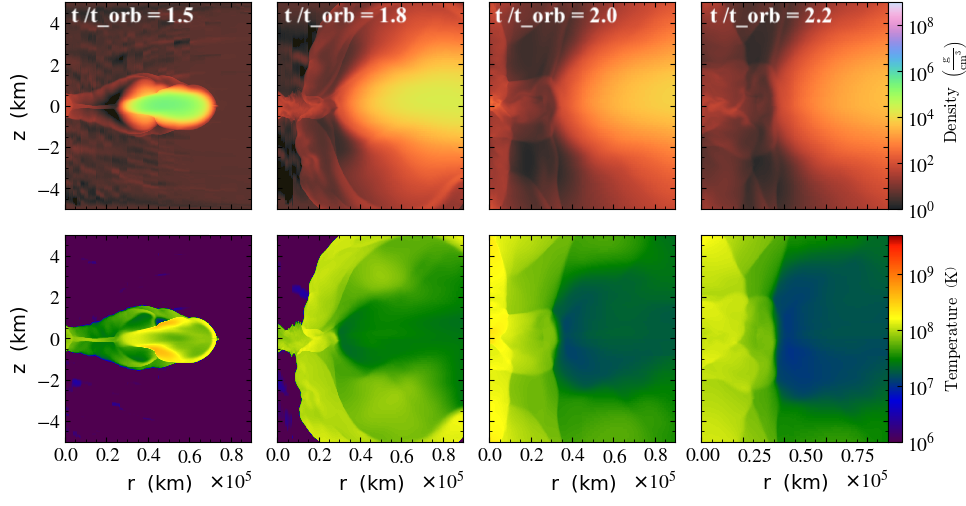}
\caption{Snapshots from the axisymmetric 2D FLASH simulation for model X$_3$ containing a $1.2\,$M$_\odot$ ONe WD + $5.0\,$M$_\odot$ BH binary. The panels show the evolution of density and temperature until $t/t_{\rm orb}=2.2\  (28.3 {\rm sec})$ for the disc approximating the remnant of the WD soon after the disruption. In all the panels, the BH accretor is located at the origin, and the panels show an R-Z region of $10^5\times 10^5\,{\rm km}$.
}
\label{fig:Model_X3_FLASH}
\end{figure*}

\begin{figure}
\includegraphics[width=\linewidth]{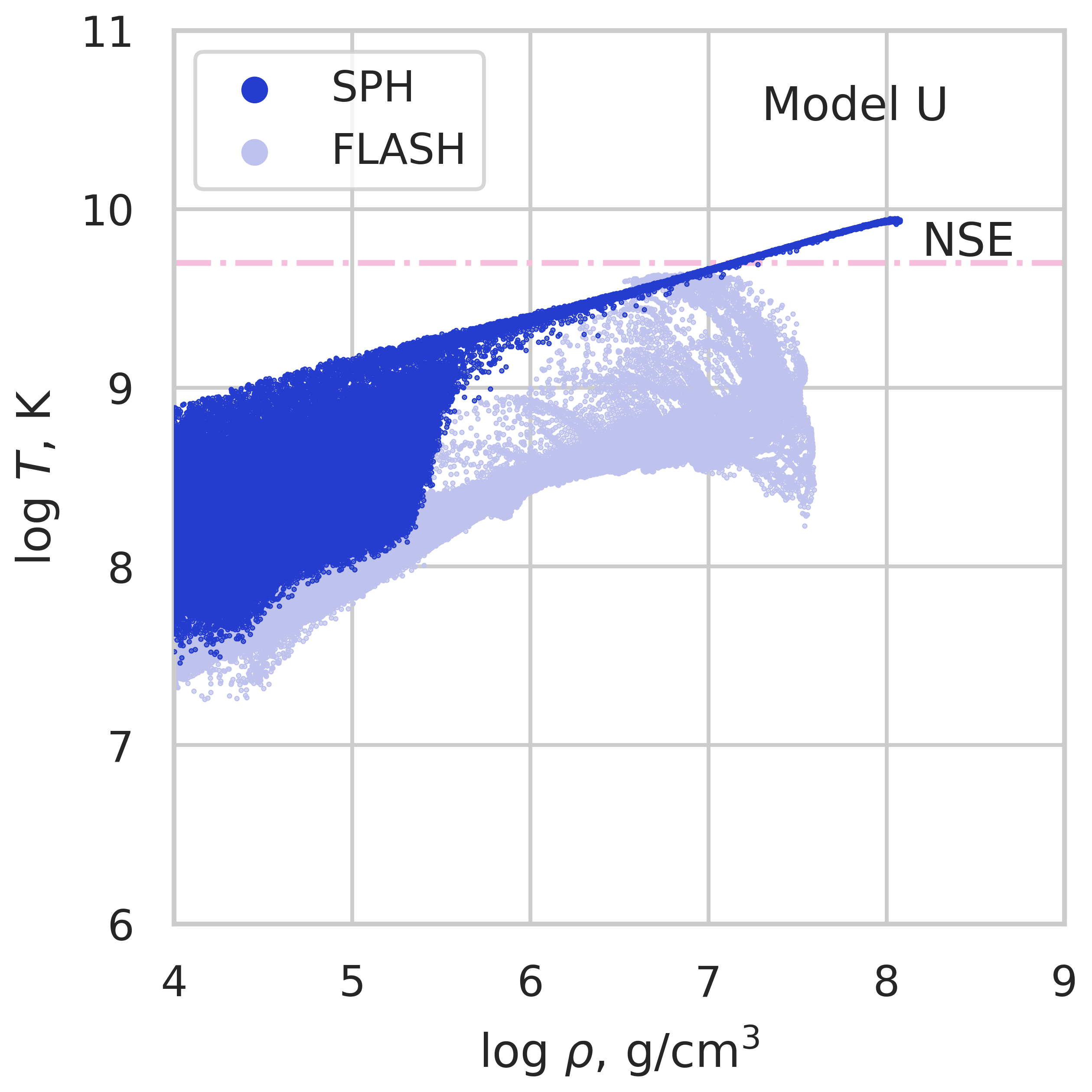}
\caption{Comparison of the density-temperature structure for model U of a $0.9\,$M$_\odot$ ONe WD + $1.4\,$M$_\odot$ NS binary, roughly at two orbital periods after the WD shredding, simulated with our 3D SPH and 2D FLASH codes (shown are SPH particles and FLASH cells, respectively). The dot-dashed line indicates the boundary for nuclear statistical equilibrium (NSE). The axisymmetric FLASH model produces a somewhat different range of densities and reaches close to the conditions for NSE, which is a possible reason why binaries with a more massive WD have been challenging to simulate with a nuclear FLASH setup.}
\label{fig:SPH_FLASHRhoT}
\end{figure}

\subsection{Comparison between the 3D and 2D models}
\label{sec:Res2D3DComparison}

In this section, we investigate the likely reasons why the mergers with massive ONe WDs have been challenging to model with earlier axisymmetric 2D simulations. We do this by comparing our 3D hydrodynamic simulations with Lagrangian SPH code Water to our 2D hydrodynamic simulations with Eulerian code FLASH.

%\review[inline]{Which models worked (Alexey)}
From the models in Table~\ref{tab:Models}, we have simulated in FLASH models U, U$_2$, both containing a $0.9\,$M$_\odot$ ONe/CO WD with a $1.4\,$M$_\odot$ NS companion, and models X$_2$ and X$_3$ with a $1.2\,$M$_\odot$ ONe WD and a massive $2.0\,$M$_\odot$ NS and a $5.0\,$M$_\odot$ BH companions, respectively. We have also verified that these models are still well-behaved in FLASH even if nuclear burning is turned on. Therefore, in particular, we conclude that the ONe composition by itself does not make FLASH simulations more challenging. We have further found that nuclear FLASH setup did not behave well for models V, W, X and Y containing WDs with masses above $1.0\,$M$_\odot$ with a $1.4\,$M$_\odot$, while non-nuclear runs did not experience such problems.

We find that the overall merger morphology agrees well with \citet{Zenati19b}. Specifically, 2D FLASH simulations start with a WD initialised as toroidal material subsequently spread viscously to represent a merger. In contrast, realistic WDs and the resulting mergers are 3-dimensional. As a result, FLASH mergers are more energetic and faster-evolving than their SPH counterparts. The converging flow at the origin produces a significantly larger amount of ejecta ($M_{\rm ej, FLASH}\approx 0.15-0.3\,{\rm M}_\odot$) compared to SPH ($M_{\rm ej, SPH}\approx 0.005-0.01\,{\rm M}_\odot$). This ejecta in FLASH might be attributed to the artificial initial setup and the convergent geometry of the flow. The post-merger discs, as in \citet{Zenati19b}, are hotter and more spread than in SPH, which may also be attributed to the initial energetics. Despite these differences, the viscous evolution in the post-merger phase in FLASH is closer to that of SPH binaries in the pre-merger phase. This better agreement is due to the axisymmetry of realistic binaries during this phase. Similarly to \citet{Zenati19b}, the FLASH nuclear yields for models U, U$_2$, X$_2$ and X$_3$ (FLASH models that completed with nuclear evolution turned on) show qualitative agreement, within 1 dex, with the SPH results.

%\review[inline]{$\rho$ and $T$ structure in the inner disc (Alexey)}
We show example snapshots from our FLASH simulation of model X$_3$ containing a $1.2\,$M$_\odot$ ONe WD and a $5.0\,$M$_\odot$ BH in Figure~\ref{fig:Model_X3_FLASH}. As we discussed earlier in \citet{Zenati19b}, $2{\rm D}$ simulations of WD-NS mergers produce discs different in morphology and structure compared to the more realistic 3-dimensional models they approximate. The difference is related to the fact that axisymmetric models can only describe an idealised geometry of WD-NS binary disruption. Due to their axisymmetric nature, $2{\rm D}$ simulations initialise WDs as tori that evolve into a disk through viscous evolution. This evolution produces an artificial convergent shock at the origin, leading to artificially enhanced outflows and additional unbound ejecta compared to 3D models. As may be seen from Figure~\ref{fig:Model_X3_FLASH}, the bulk of the disc after the convergent shock phase is geometrically thicker than the SPH disc in Figure~\ref{fig:SPH}. Before the convergent shock phase, the disc is comparably thin to the SPH disc but is significantly colder because it has not been heated up by the shocks yet. In the 3D case, shocks occur continuously throughout the WD disruption from the start. Nevertheless, these differences do not seem to make FLASH simulations significantly more challenging for higher WD masses. While more massive WDs do lead to more energetic mergers, the FLASH disc properties in the ONe WD-NS simulations are relatively similar to the ones we observed in \citet{Zenati19b} for less massive CO WDs. Similarly, as in \citet{Zenati19b}, we also find that the nuclear energy release is small in our models compared to the gravitational binding energy of SPH particles. Therefore, nuclear evolution in the binaries with a massive WD donor has a similar negligible dynamical importance as in less massive CO WDs.

%\review[inline]{Nuclear burning and NSE (Alexey)}
We find that the reason why ONe WD-NS binaries have been challenging in earlier axisymmetric simulations is likely related to the fact that massive WDs lead to nuclear statistical equilibrium (NSE) conditions in the disc. Modelling such conditions is computationally more difficult, e.g. \citet{Timmes1999,Timmes2000b}. Indeed, as we discuss in Section~\ref{sec:ResNuc}, more massive WD mergers are more energetic, lead to higher temperatures in the disc and, for that reason, produce more nuclear ashes. In Figure~\ref{fig:SPH_FLASHRhoT}, we show the comparison for density-temperature structure for model U at about two periods after the merger (i.e. after the start of the simulation, in the case of FLASH), for the SPH particles and FLASH cells\footnote{It should be noted that FLASH points in Figure~\ref{fig:SPH_FLASHRhoT} represent grid cells rather than equal masses of WD material, as opposed to the SPH points in the figure.}, superimposed with the characteristic boundary for NSE \citep{Iliadis2007}. FLASH disc is shown slightly before the convergent shock phase and is therefore still relatively cold. We see that at this point, FLASH models with a $0.9\,$M$_\odot$ WD reach temperatures in close proximity to the NSE boundary. Furthermore, FLASH models with more massive WDs cross it. Once this happens, FLASH simulations become prohibitively slow if nuclear evolution is enabled. At the same time, more massive NS or BH companions in models X$_2$ and X$_3$ lead to initially wider orbits and overall lower disc temperatures.

%\review[inline]{NSE in detail, comparison to \citet{Fernandez19}  (Alexey)}
NSE is challenging to simulate because the timescales for nuclear reactions, in this case, become vanishingly small. For example, as may be seen from Figure~\ref{fig:SPH_FLASHRhoT}, non-nuclear SPH simulations cross the NSE boundary, with the upper-right slope on the figure formed by the particles near the NS. The slope itself follows approximately $T\sim \rho^{1/3}$ dependency, expected for adiabatically compressed radiative pressure-dominated gas. When we applied the MESA PPN nuclear solver \citep{Paxton2011} to the SPH tracks that reach NSE, we found that the solver becomes computationally inefficient due to significantly decreased timesteps. In contrast, the nuclear Torch code, which uses dedicated numerical methods to treat NSE \citep{Timmes1999,Timmes2000b}, performs the same computation several orders of magnitude faster. In the \citet{Fernandez19} study, for example, the authors use a lightweight nuclear scheme, likely not designed to treat NSE specifically. Therefore, similarly to our FLASH setup, such models have likely been computationally prohibitive, requiring too small timesteps to resolve the nuclear evolution.

\begin{table*}
    \centering
    \begin{tabular}{|c|c|c|c|c|c|c|c|c|c|c|}
    \hline
         Transient & Peak mag & Colour & Width & Slope & Spectra & $^{56}$Ni & $v_{\rm phot}$ & Rate & Comment & Reference \\
    \hline
    AT2018kzr & Ok  & Ok & Good & Good & Ok & Good & Good & -- & ONe WD-BH & \citep{Gillanders2020} \\
    Faint Iax & Good & Ok & Good & Good & Good & Good & Good & -- & Faint end & \citep{Jha2017}\\
    Faint Ic & Good & Bad & Ok & Bad & Bad & Good & Good & Ok & -- & \citep{Barbarino2020} \\
    Faint 1991bg & Bad & Bad & Bad & Ok & Bad & Bad & Good & -- & -- & \citep{Doull2011} \\
    Faint Ca-rich & Bad & Bad & Bad & Bad & Bad & Ok & Good & Ok & -- & \citep{De2020} \\
    Faint red & Bad & Good & Bad & Bad & Bad & Possible & Bad & Bad & -- & \citep{Drout2014} \\
    Kilonovae & Good & Bad & Bad & Bad & Bad & Ok & Bad & Ok & -- & \citep{McBrien2021} \\
    LRN'e/ILRT's & Good & Good & Good & Good & Bad & Bad & Good & -- & -- & \citep{Williams2020} \\
    % & & & & & & & & & & \\
    \hline
    \end{tabular}
    \caption{Comparison table of known faint peculiar red transients to the properties predicted by our models. We compare our models to the candidate ONe WD disruption event AT 2018kzr, faint end of Iax, Ic, 1991bg and Ca-rich supernovae, faint red transients, kilonovae, luminous red novae (LRNe) and intermediate-luminosity red transients (ILRTs). For comparison, we consider lightcurve properties (colour, peak magnitudes, decline slope, peak width), spectral properties (spectra, amount of nickel-56, photometric velocities) and the expected formation rates. Based on the comparison, transient AT2018kzr was likely produced by an ONe WD-BH merger, while CO and ONe WD-NS mergers most likely produce faint type Iax supernovae.}
    \label{tab:Transients}
\end{table*}

\section{Discussion}
\label{sec:Disc}

%\review[inline]{Short summary (Alexey)}
In this study, we have constructed the first lightcurves and spectra for ONe WD-NS/BH mergers and also have conducted the first such simulations with a 3D hydrodynamics code. Generally, we predict that ONe WD-NS mergers should typically produce month(s)-long red/infrared transients potentially occupying the faint end of existing nuclear-powered transients, while ONe WD-BH mergers should produce week(s)-long transients. In this section, we review the likely observational counterparts and other observational manifestations for these systems and summarize the current modelling status for WD-NS/BH binaries in general.

\subsection{How the modelling uncertainties affect the observations}

%\review[inline]{Main modelling uncertainties, effect of ejection velocity (Alexey)}
There are three main uncertainties in the present-day models of WD-NS/BH mergers. All of them result from our lack of knowledge of how exactly WD-NS/BH binaries eject the WD material after the merger. One key uncertainty is the ejection velocity, which may range between a few $100\,{\textrm km}/{\textrm s}$ and several $1000\,{\textrm km}/{\textrm s}$. The velocity has a direct effect on the duration and brightness of the transients. Typical fastest-velocity counterparts will produce transients reaching bolometric magnitudes of up to $-16.5$ and lasting for about a month. In contrast, slower-velocity models produce one magnitude fainter transients lasting for several months. Observations of photospheric expansion velocities for such transients will help to constrain the models.

%\review[inline]{Ejected mass fraction (Alexey)}
The other key uncertainty is related to the total amount of the WD mass ejected after the merger. Our SPH simulations show that the amount of material ejected dynamically during the mergers is very small, of the order of a fraction of a per cent of the WD mass. However, in our fiducial lightcurve modelling, we assume that most of the WD mass is ejected later by some feedback process, which is certainly an optimistic assumption. In principle, the fraction of ejected mass may indeed be close to a hundred per cent since the energy from accreting about $0.01\,$M$_\odot$ onto the NS, at accretion efficiency of $15$ per cent is sufficient to unbind the rest of the material. However, this reasoning would be valid only if all the released energy were deposited to the WD material rather than lost from the system. In reality, the mass accreted by the NS is likely larger since some energy will get lost through a collimated jetted outflow. As we saw from the reduced-mass models, smaller fractions of ejected mass produce fainter, faster-evolving, but otherwise similar transients. The fraction of the ejected mass in real systems may likely be constrained by the observations of the brightest transients of the class. 

%\review[inline]{Composition of the ejecta (Alexey)}
The last key uncertainty is related to the composition of the ejected material, both in terms of whether it is homogeneously mixed and whether all the nuclear ashes get ejected from the system. The degree of homogeneity may likely be assessed by the smoothness of the lightcurves and from the spectra, as we discuss in Section~\ref{sec:ResTransients}. The fraction of the ejected nuclear ashes will be hopefully possible to analyse through spectra. In particular, the amount of the ejected $^{56}$Ni may have a significant impact both on the lightcurve and the spectra. In contrast, the initial composition of the WDs, which may vary in relative fractions of C, O, Ne and Mg, does not have a strong effect on the final lightcurves and spectra. This conclusion follows from the overall similarity of lightcurves and spectra of U and U$_2$ models that differ only in their initial compositions.

If the actual disc viscosity is significantly larger than the numerical SPH viscosity ($\alpha_{\rm SPH}\approx 0.01$) and if at later times viscous spreading of the disc delivers additional unburnt WD material close to the NS before the feedback processes eject it, one may expect that additional $^{56}{\rm Ni}$ and intermediate-group elements get produced. Furthermore, if such material gets further ejected, its effect on the optical transients resulting from WD-NS mergers would be degenerate with the WD mass. In other words, the transients from our current models with lower-mass WDs would observationally become similar to the transients from our current models with higher-mass WDs, as the latter produce more $^{56}{\rm Ni}$. The assumptions needed for such a scenario may be tested with detailed future radiative-MHD simulations.

%\review[inline]{Composition of the ejecta (Alexey)}
Quite possibly, realistic ejecta is asymmetric and multi-component. The low-mass dynamical ejecta lost in the early stages of the merger are mostly planar, the possible fast outflow driven by the NS is likely conical, and the slow and later ejecta resulting from disc dispersal can be either spherical or planar. Each component of the ejecta would contribute to the LCs and spectra. The resulting LCs and spectra will probably be a superposition of the models presented in Figure~\ref{fig:ModelComparison}, with an earlier faster and bluer component followed by a later slower red/infrared component.

%\review[inline]{Current modelling (Alexey)}
In this study, we have argued that 2D axisymmetric codes should, in principle, be capable of simulating all ONe WD-NS mergers, provided that the nuclear solver can efficiently treat the nuclear statistical equilibrium phase, occurring in the binaries with a massive WD. As we discuss in \citet{Zenati19b}, 3D simulations, as performed here, are generally expected to produce geometrically realistic mergers unattainable in axisymmetry. The convergent character of the disc in axisymmetric simulations leads to energy deposition near the origin, resulting in artificial outflows. In contrast, 3D simulations show that the material immediately after the merger is mostly gravitationally bound.

%\review[inline]{Future modelling (Alexey)}
Future modelling of WD-NS binaries should likely be performed in 3D, extending into the later phase when the material gets ejected. In this regime, accurately modelling accretion onto the NS and the resulting feedback becomes important, which may be done easier by elaborating on or coupling to existing 1D/2D studies, e.g. \citet{Margalit16, Fernandez19}. Indeed, as may be seen in Fig.~\ref{fig:SPH}, the disrupted WD becomes fully axisymmetric and Keplerian. The 1D/2D models can be realistically initialised at this stage instead of WD-NS binaries before the disruption. Such simulations can be used to model how the material is eventually ejected from the binary. In particular, radiation transfer and subgrid energy injection from near the accretor would be necessary. Additionally, including magnetohydrodynamic effects in future studies may be needed to accurately capture disc evolution and material ejection. Assessing the asphericity of the ejecta, see e.g. \citet{Korobkin2020}, may also have important implications for the observations.

\subsection{Likely optical transient counterparts}

%\review[inline]{Short summary (Alexey)}
The first transient from a WD-NS binary will likely have been originated from an ONe WD-NS binary. The odds favour these progenitors because they produce transients with bolometric magnitudes of typically about -15.5, about 1.5 magnitudes brighter, and also about two times longer than the transients from typical CO WD-NS binaries. Additionally, both ONe and CO WD-NS binaries have at least comparable inspiral rates, as discussed in Section~1. The relatively large amount of $^{56}$Ni produced in ONe WD-NS mergers, up to about $0.05$~--~$0.1\,$M$_\odot$, makes them potentially occupy an ultrafaint end of type Iax-like nuclear supernovae (2008ha-like SNe) or a faint end of supernovae type Ic, depending on whether the Si lines are pronounced in the spectra, e.g. \citet{Jha2019, Perley2020}. ONe WD-BH mergers are expected to produce a separate type of distinct shorter, week(s)-long, transients.

%\review[inline]{Expected detection rates (Alexey)}
Moreover, the current all-sky synoptic surveys likely have already detected the transients from ONe WD-NS mergers. To obtain the expected detection rates, we use the values for model W, which represents the Galactic WD-NS binary PSR J1141-6545 and is a typical representative of ONe WD-NS populations. The peak AB r-band magnitudes for model W are between $-15$ and $-16.25$ depending on the mass ejection scenario, as may be seen from Figure~\ref{fig:ModelComparison}. Furthermore, the lightcurve fades negligibly, by at most $0.15$ mag, within two weeks from the peak. Given the peak  AB r-band sensitivity of $19.7$ and $21.0$ for the currently operating ATLAS \citep{Tonry2018} and ZTF \citep{Bellm2019} surveys, respectively, they are able to observe such transients for two weeks after the AB r-band peak from distances up to $90$~--~$240\,{\rm Mpc}$ and $160$~--~$260\,{\rm Mpc}$, respectively. Furthermore, we assume ONe WD-NS merger rate of $200\,{\rm Myr}^{-1}$ per MW-like Galaxy, e.g. \citep{Bobrick2017, Toonen2018}, use the local universe and Galactic B-band luminosities of $1.98\cdot 10^8\,L_{{\rm B},\odot}/{\rm Mpc}^3$ \citep{Kopparapu2008} and $1.7\cdot 10^{10}\,L_{{\rm B},\odot}$ \citep{Kalogera2001}, respectively, and assume that the ONe WD-NS merger rate follows the blue luminosity. We find the expected detection rates for the ongoing ALTAS and ZTF surveys of $1.6$~--~$7.4\,{\rm yr}^{-1}$ and $9.7$~--~$44.4\,{\rm yr}^{-1}$, respectively. Similarly, the upcoming Vera Rubin Observatory (LSST) will have an AB r-band detection limit of $24.7$ \citep{LSST2009}, and it will pick up these transients at a rate of $1600$~--~$7400\,{\rm yr}^{-1}$ from distances of $870$~--~$1400\,{\rm Mpc}$. Therefore, present surveys are able to detect these transients and thereby support or constrain these models and such detections should be much more common in the near future.

%\review[inline]{Specific expectations from the modelling (Alexey)}
We expect ONe WD-NS mergers to produce red/infrared transients with peak AB r magnitudes of $-14.5$ to $-16.75$, between one and a few months in duration (with fainter transients lasting longer). We expect these transients to produce up to $0.1\,$M$_\odot$ of $^{56}$Ni, have photometric velocities between $500$ and $10000\,{\rm km}/{\rm s}$ and have lightcurves and spectra resembling the ones from our models. The lightcurves might look like combinations of those from fast and slow ejecta models shown in Figure~\ref{fig:ModelComparison}. The ejecta mass may reach up to about $1.3\,{\rm M}_\odot$ but may also be lower. About $0.02$~--~$0.08\,$M$_\odot$ of helium and up to $0.02\,$M$_\odot$ of hydrogen should be present in the ejecta, although the spectroscopic signatures may be too faint for detection. Although NS could receive a kick at birth, these are not likely to significantly affect the galactocentric offsets of these transients from the hosts (see discussion in \cite{Perets21}). They may either be biased to late-type hosts or be present in both early and late-type hosts. Both possibilities are allowed because the delay time distribution for these systems may be relatively short or extend to the Hubble time depending on the population model parameters \citep{Toonen2018}.

%\review[inline]{SN AT 2018kzr - candidate ONe WD-BH (Alexey)}
We now consider the known classes of observed faint and red transients and compare them to our expectations from ONe WD-NS/BH binaries. We summarise the comparison in Table~\ref{tab:Transients}, starting with the most promising AT2018kzr-like events. Supernova AT 2018kzr has been earlier proposed to originate from an ONe WD-NS/BH merger \citep{McBrien2019,Gillanders2020}. We see that their bright peak bolometric luminosities of $10^{43}\,{\rm erg}/{\rm s}$ exceed values reached by WD-NS/BH mergers. The initially blue colours, short decay times and expansion velocities seen in AT2018kzr are also not typical for WD-NS/BH mergers. The main peak in our models occurs at lower luminosities. However, for ONe WD-BH mergers, it might not be well covered by observations. We find that only by initialising X$_3$ model with $10^5\,{\rm K}$ hot material at $0.5$ days post-merger may produce the early peak similar to the one observed in SN AT2018kzr. Such a model could correspond to fallback luminosity \citep{Dexter2013} heating the ejecta from inside in addition to radioactive decay. The early spectra then show a reasonable agreement with model X$_3$ initialised this way, but the inferred elemental abundances agree only partly. In particular, the fractions of $^{56}$Ni and $^{16}$O are reasonably consistent with our model, but the fractions of Fe, Ti, Si and Cr all disagree. Altogether, these agreements strengthen the case that SN 2018kzr might be an outcome of an ONe WD-BH merger only assuming additional physical heating mechanisms for the ejecta while the ONe WD-NS merger scenario is unlikely.

%\review[inline]{Faint end of Iax's - candidate CO/ONe WD-NS (Alexey)}

Peak brightnesses of faint type Iax supernovae could match the brightness reached by ONe WD-NS mergers. Bright type Iax explosions are believed to be produced by partial deflagrations of WD's \citep{Jordan12}, see also \cite{Kromer13,Foley2013}, leaving behind bright bloated exotic WD remnants, e.g. SN 2012Z \citep{McCully2014}. However, the observed type Iax SNe span almost five magnitudes in brightness (2 orders in luminosity) and may well enclose multiple classes of physical events \citep{Jha2017}. Indeed, faint 2008ha-like type Iax supernovae span an attractively wide range of peak luminosities, down to $10^{40.7}\,{\rm erg}/{\rm s}$ at peak \citep{Srivastav2020}. In our fiducial model, this range covers the transients from ONe WD-NS binaries ($10^{41.5}$~--~$10^{42}\,{\rm erg}/{\rm s}$), the fainter transients from CO WD-NS binaries ($10^{41.1}$~--~$10^{41.5}\,{\rm erg}/{\rm s}$) \citep{Zenati19b} and yet fainter transients expected from lower-mass He WD-NS binaries ($10^{40.7}$~--~$10^{41.1}\,{\rm erg}/{\rm s}$)\footnote{Simulations of He WD-NS binaries will be published in a follow-up study}. The durations of faint Iax supernovae are comparable to our models and increase with brightness, as also seen for our models. The colours of the faint Iax supernovae are typically too white near the peak (B-V$\approx$V-R$\approx$0) but otherwise match our lightcurves. The estimated amount of $^{56}$Ni and the photospheric velocities for these transients match our fast-ejecta scenario. Finally, spectra from some of the transients after 20 days past peak (e.g. 2005hk, 2002cx, 2008ha, 1998bu) also resemble the spectra of our transients at a similar epoch. Therefore, our study suggests that the faint end of type Iax supernovae is occupied by the transients from CO/ONe WD-NS mergers. If this interpretation is correct, the observations of faint type Iax supernovae support the fast-ejection scenario. The colour and spectra in the first days of the transient would then suggest that an additional mechanism, such as shock-powered circumstellar interaction \citep{Margalit16} or fallback luminosity \citep{Dexter2013}, might be heating up the ejecta during that time. Finally, we note that the leading, failed-detonation/partial deflagration model for Iax SNe \cite{Jordan12,Kromer13}, well explains most Iax SNe but fail to explain the faint end, possibly providing a clue that these fainter SNe may indeed have a different origin.

%\review[inline]{Faint end of Ic's - unlikely ONe WD-NS (Alexey)}
Some ultrafaint type Ic supernovae may also be from WD-NS mergers in origin, given that Si lines are not strongly pronounced in their spectra. The faintest members of the iPTF type Ic sample \citep{Barbarino2020}, for example, iPTF 13djf, reach peak bolometric magnitudes of $-15$, consistent with our ONe WD-NS models. However, such supernovae are not as red (g-r$\approx$1) as our transients. Faint type Ic supernovae also fade faster, by about one magnitude over two weeks. They also do not show a correlation between the fading time and the peak luminosity observed in ONe WD-NS mergers \citep{Barbarino2020}. Also, while the inferred $^{56}$Ni amount and ejecta mass may be consistent with these transients, their spectra for the faintest objects are often either not available or qualitatively disagree with our spectra, e.g. \citep{Yao2020}. Therefore, there does not seem to be a strong match between ultrafaint type Ic supernovae and ONe WD-NS transients, although it may not be firmly excluded either. 

%\review[inline]{Faint end of 1991bgs, Ca-rich or rapid red transients - unlikely ONe WD-NS (Alexey)}
The faintest members of several other types of peculiar transients, for example, 1991bg-like type Ia supernovae \citep{Filippenko1992, Doull2011}, Ca-rich/strong transients \citep{Perets2010, De2020}, or faint red transients \citep{Drout2014, Margalit16}, may in principle approach the peak magnitudes of the most massive of our ONe WD-NS models. However, we would then also expect fainter members of these classes, produced by lower-mass WD-NS binaries, to have also been observed. Furthermore, the faintest members of these classes also seem to differ from our models. All known 1991bg-like supernovae, for example, are different in colour and spectra and fade faster than model Y, e.g. \citet{Taubenberger2008}. Additionally, 1991bg-like supernovae require more than $0.07\,$M$_\odot$ of $^{56}$Ni \citep{Mazzali1997}, which is near the limit of our models. Similarly, Ca-strong transients are all brighter than our WD-NS models and differ in colour, show a faster decline and most importantly, our models do not show strong calcium emission at late times. Recently, \citet{Polin2021} showed that, due to non-LTE effects, as little as $0.02\,{\rm M}_\odot$ of $^{40}$Ca can produce the Ca/O emission-line ratios observed in Ca-rich transients. However, our models produce only a few times $0.001\,{\rm M}_\odot$ of $^{40}$Ca and much more oxygen from the unprocessed material compared to \citet{Polin2021}, thus making the appearance of significant Ca/O ratios very unlikely. And also similarly, faint red transients are all faster-evolved and brighter than the brightest transients that ONe WD-NS binaries produce in all our models. While being red overall, the blue part of the spectra for faint red transients differs significantly from our models. They also seem to imply too high ejecta velocities, and their formation rates exceed the expected value for ONe WD-NS binaries by a factor of 5 to 8. Therefore, we do not find any significant evidence for a connection between ONe WD-NS mergers and 1991bg-like or Ca-rich supernovae or faint red transients.

%\review[inline]{Ultrafaint red: Kilonovae, Sprites, ILRTs, LRNe (Alexey)}
There are also several yet fainter classes of red transients, although we find that they cannot serve as counterparts to WD-NS mergers. For example, kilonovae are unlikely to be associated with WD-NS mergers. Despite having comparable peak magnitudes to some of the models, kilonovae are associated with gravitational wave sources \citep{Arcavi2017}, show much faster evolution, too large photospheric velocities and very different spectra \citep{McBrien2021}. Similarly, while luminous red novae (LRNe) and intermediate luminosity red transients (ILRT) show potentially similar photometric evolution to our slow-ejecta scenario, their spectra show strong hydrogen emission lines which are not possible for our models \citep{Williams2020}.

\subsection{Other observational manifestations}

Apart from the transients they produce, WD-NS/BH binary mergers have a variety of other observational implications. Here we list the most important ones in connection to this study.

%\review[inline]{Galaxy enrichment general (Alexey)}
WD-NS mergers serve as a separate potentially important source of Galactic chemical enrichment. As we showed in Section \ref{sec:Intro}, massive WD-NS mergers dominate the enrichment of the Galaxy compared to all WD-NS mergers. They spiral in at $6$~--~$20$ per cent of type Ia rate. Since WD-NS binaries are formed from more massive stars, their delay time distribution may be shorter than for typical type Ia supernova progenitors, e.g. \citet{Toonen2018}, which would be further supported if the type Iax interpretation for WD-NS mergers is correct. Furthermore, since WD-NS mergers produce about $6$ times less Fe than type Ia supernovae, their effect may be relatively easily disentangled from type Ia contributions. 

%\review[inline]{Mn55 contribution (Alexey)}
The early production of $^{55}$Mn in the Galaxy, for example, may potentially have been significantly affected by WD-NS mergers. By using model W in Table~\ref{tab:Nucyields}, which is representative of all WD-NS mergers, and by adopting the solar elemental mass ratio Fe/Mn=$119\pm15$ from \citet{Grevesse2010}, we find that WD-NS binaries enrich the Galaxy with an abundance ratio of $[{\rm Mn}/{\rm Fe}]=0.57$. Since this is $0.3$ to $0.8$ dex larger than expected for type Ia supernovae \citep{ Seitenzahl2013,Seitenzahl2013b}, we expect WD-NS binaries to have produced between $2$ and $22$ per cent of the current Galactic $^{55}$Mn, depending on the uncertainties in the WD-NS inspiral rates and the yields of type Ia supernovae. Similarly, depending on the delay time distributions, the manganese abundances in the early Galaxy at $[{\rm Fe}/{\rm H}]=-1$ may also have been affected significantly \citep{Seitenzahl2013b}. However, reproducing the low $[{\rm Mn}/{\rm Fe}]=-0.5$ abundances in the early Galaxy still requires an additional polluter scenario, such as sub-Chandrasekhar metallicity-dependent type Ia explosions \citep{Seitenzahl2013b,Gronow2021}.

%\review[inline]{Galaxy, other elements (Alexey)}
We also expect that WD-NS mergers may be responsible for $1$ to $3$ per cent of iron in the Galaxy.  The enrichment fraction in carbon, oxygen and neon from WD-NS mergers may be up to ten times larger than by other elements (accounting for both CO and ONe WD-NS mergers). However, AGB stars likely dominate C production, while core-collapse supernovae likely dominate O and Ne production in the Galaxy \citep{Johnson2019}. Similarly, type Ia supernovae likely dominate alpha-element production over WD-NS binaries, e.g. \citet{Lach2020}. On the other hand, the production of short-lived $^{53}$Mn per one ONe WD-NS merger may exceed that of type Ia supernovae by more than ten times, e.g. \citet{Fink2014}, and therefore can make an important contribution to the Galaxy. $^{53}$Mn decays through electron capture with half-life of $4\,$Myr, produces no significant gamma-emission \citep{Dressler2012} and hence cannot be observed in X-ray. However, ONe WD-NS mergers can certainly contribute to the Earth enrichment by $^{53}$Mn from type Ia supernovae if it takes place \citep{Schaefer2006}. Finally, we may speculate, if the WD material manages to escape from the direct vicinity of the NS through an outflow, WD-NS binaries may potentially contribute to the Galactic r-process enrichment.

%\review[inline]{Other signatures of the merger (Alexey)}
The merger may also be observed as LISA \citep{LorenAguilar2005} or DECIGO \citep{Kinugawa2019} GW transients. Owing to their higher masses and more compact orbits, the detection rate will be comparable to that of DWD inspirals despite a lower formation rate. Before the merger, one might expect a bright X-ray transient at the location of the event, e.g. \citet{Bobrick2017}. Such X-ray emission will be induced by the accretion and will last for hours to days before the merger, with the duration of the X-ray phase allowing to put constraints on the mass of the WD. During the merger, apart from the transient, one may also expect a jetted outflow from the accretor. Hence, depending on the viewing angle, we may expect a UV or X-ray emission accompanying the transient in the first hours to days after the transient, as well as a shock emission in the optical bands \citep{Fernandez19}. In the days to months after the merger, the fallback accretion may lead to a new temporary X-ray source at the location of the merger \citep{Dexter2013}. Its luminosity would be potentially comparable to that from known ultra-luminous X-ray sources. Detection of such a post-transient source could help identify these systems. The optical observation of a luminous red remnant at the site of the faint SN Iax explosion SN 2008ha \citep{Foley2014} potentially originating from CO/ONe WD-NS mergers may provide essential clues on the evolution of their remnants. Additionally, a remnant hot X-ray bright potentially kicked NS would be expected from after the explosion. Potentially magnetised by the merger, the NS remnants may be observed as Fast Radio Burst (FRB) sources \citep{Petroff2019}. This outcome is significant for FRBs observed in globular clusters \citep{Kirsten2021, Katz2021} where magnetars cannot form naturally from massive stars. In the long term, the disc from WD remnants around the NS/BH produce pulsar or black hole planets \citep{Margalit17}. Alternatively, if the NS failed to eject a large part of its material, it might collapse into a black hole. Therefore, observations of a NS remnant would allow one to put constraints on the amount of mass ejected during the merger. 

%\review[inline]{SNR connection (Alexey)}
Remnants from the CO/ONe WD-NS mergers should leave behind oxygen-rich nebulae. This class of nebulae displays prominent oxygen emission and a lack of hydrogen lines. Currently, 8 such nebulae are known: Cas A \citep{Kirshner1977}, Puppis A \citep{Winkler1985} and G292+1.8 \citep{Goss1979} in the Galaxy, N132D \citep{Danziger1976} and 0540-69.3 \citep{Mathewson1980} in the LMC, and 1E 0102.2-7219 \citep{Dopita1979}, 0103-72.6 \citep{Park2003} and B0049-73.6 \citep{Hendrick2005} in the SMC. Such nebulae have been traditionally associated with remnants from stripped core-collapse supernovae. However, in our Galaxy, one WD-NS merger occurs every $5000$ years. Therefore, the youngest nebula from such mergers will likely be a few thousand years old, and there should be several such nebulae in the Galaxy and its satellites. From the known oxygen nebulae, the 5-10\,{\rm kyr}-old Puppis A, 1.5\,{\rm kyr}-old  G292+1.8, 2.5\,{\rm kyr}-old N132D, 1\,{\rm kyr}-old 1E 0102.2-7219, 10\,{\rm kyr}-old 0103-72.6 and 14\,{\rm kyr}-old B0049-73.6 show prominent neon emission and almost complete lack of hydrogen. Furthermore, for example, Cas A and SNR N132D have a planar toroidal morphology \citep{Law2020}, which a binary WD-NS merger may produce. Taken together, some of the oxygen nebulae may quite probably be of WD-NS origin. Detailed abundance comparisons as, e.g. in \citet{Zhou2021}, must be used to establish a firm connection to progenitors. Conversely, the nebulae identified with WD-NS mergers will give new insights into the merger mechanisms.

\section{Summary}

%\review[inline]{Summary for WD-NS simulations (Alexey)} 
In this study, we have performed the first hydrodynamic simulations of ONe WD-NS mergers and the first 3-dimensional simulations of ONe WD-NS/BH mergers. By calculating the nuclear evolution of material in such mergers, we obtained the expected yields which such mergers can produce. While the fraction of ejected material is uncertain, we expect that ONe WD-NS binaries will be the main polluter of the environment compared to all WD-NS/BH binaries. In particular, such transients might have a significant role in the production of $^{55}$Mn in galaxies. We have also synthesised the first lightcurves and spectra for ONe WD-NS/BH mergers. While the details of how the material is ejected from the system after the merger are uncertain, we are now able to test the possible scenarios against the observations.

%\review[inline]{Summary for observations (Alexey)}
On the observational side, we expect that the transients from ONe WD-NS mergers very likely have already been detected by the ongoing all-sky surveys. The brightness of ONe WD-NS binaries is related to the fact that they produce a significant amount of $^{56}$Ni, ranging from $0.05\,{\rm M}_\odot$ in the typical cases and reaching up to $0.1\,{\rm M}_\odot$ for the massive WDs. The resulting transients in all scenarios are expected to be red/infrared, evolving on month(s) timescales and having bolometric magnitudes of up to $-16.5$. By comparing the lightcurves and spectra from our systems to the observed events, we find the most natural counterparts to CO and ONe WD-NS mergers are the faint type Iax supernovae. At the same the recent ONe WD disruption candidate event SN AT2018kzr was potentially produced by an ONe WD-BH merger, but only assuming an additional energy source apart from the radioactive decay. WD-NS mergers are a one-parameter family of systems that depend mostly on the WD mass, while WD-BH mergers have a relatively small range of WD/BH masses allowed for them to merge \citep{Church2017}. If the association is indeed correct, detailed comparisons to the observed transients may provide a wealth of new constraints on the models. Therefore, these systems may pave the way towards a complete identification of evolutionary models of compact binary mergers and the transients they produce.

\section*{Acknowledgements}

We thank Silvia Toonen, Ashley Ruiter, Thomas Tauris, Christian Setzer, Josefin Larsson, James Gillanders, Stephen J. Smartt, Jennifer Johnson, Kojiro Kawana, Paul Callanan, Stephen Justham, Noam Soker, Francis Timmes, Daan van Rossum and the Anonymous Referee for their valuable comments and discussions at different stages of this study. The simulations (other than FLASH simulations) were performed on the resources provided by the Swedish National Infrastructure for Computing (SNIC) at the Lunarc cluster. FLASH simulations were performed on the Astric computer cluster of the Israeli I-CORE center. We acknowledge support for this project from the European Union's Horizon 2020 research and innovation program under grant agreement No 865932-ERC-SNeX.

%%%%%%%%%%%%%%%%%%%%%%%%%%%%%%%%%%%%%%%%%%%%%%%%%%

\section*{Data availability}

The data underlying this article will be shared on reasonable request
to the corresponding author.

\bibliographystyle{mnras}
\bibliography{ONeLits}
%%%%%%%%%%%%%%%%%%%%%%%%%%%%%%%%%%%%%%%%%%%%%%%%%%

%%%%%%%%%%%%%%%%% APPENDICES %%%%%%%%%%%%%%%%%%%%%%

% Don't change these lines
\bsp	% typesetting comment
\label{lastpage}

\end{document}